\documentclass[11pt]{article}

\usepackage{amstext,amsgen,latexsym,amsmath}
\usepackage{amstext,amssymb,amsfonts,latexsym}
\usepackage{theorem} \usepackage{pifont}
\usepackage[dvips]{graphics,epsfig}

 \newcommand{\bs}{\bigskip} \newcommand{\ms}{\medskip}
 \newcommand{\n}{\noindent} 
 \newcommand{\hs}[1]{\hspace*{ #1 mm}}

 \def\bbox{\vrule height6pt width6pt depth1pt}

\theoremstyle{plain}
\theoremheaderfont{\bfseries}
 \newtheorem{theorem}{Theorem}[section] \newtheorem{lemma}[theorem]{Lemma}

 \newtheorem{claim}{Claim}

 \newenvironment{proof}{\par \noindent
            {\bf Proof. \hs{2}}}{\hfill$\Box$ \vspace*{3mm}}
 
 \newenvironment{proofof}[1]{\vspace*{5mm} \par \noindent
         {\bf Proof of #1.\hs{2}}}{\hfill$\Box$ \vspace*{3mm}}

\newcommand{\ignore}[1]{}

\begin{document}
%%%%%%%%%%%%%%%%%%
\pagestyle{plain}

\begin{center}
{\Large {\bf Quantum Network Coding}}
\bs\\

%\begin{tabular}{c@{\hspace{30mm}}c}
\begin{center}
{\sc Masahito Hayashi}$^1$\footnote{Also, Superrobust Computation Project, Information Science and Technology Strategic Core (21st Century COE by MEXT), Graduate School of Information Science and Technology, The University of Tokyo
}
 \hspace{10mm} {\sc Kazuo Iwama}$^2$\footnote{Supported in part by Scientific Research Grant, Ministry of Japan, 16092101.} \hspace{10mm} {\sc Harumichi Nishimura}$^3$\footnote{Supported in part by Scientifis Research Grant, 
Ministry of Japan, 18244210.}
\\{\sc Rudy Raymond}$^4$ \hspace{30mm} {\sc Shigeru Yamashita}$^5$\footnote{Supported in part by Scientific Research Grant, Ministry of Japan, 16092218.} 
\end{center}

$^1${ERATO-SORST Quantum Computation and Information Project,\\ Japan Science and Technology Agency}

{\tt masahito@qci.jst.go.jp}

$^2${School of Informatics, Kyoto University}

{\tt iwama@kuis.kyoto-u.ac.jp} 

$^3${School of Science, Osaka Prefecture University}

{\tt hnishimura@mi.s.osakafu-u.ac.jp}

$^4${Tokyo Research Laboratory, IBM Japan}

{\tt raymond@jp.ibm.com}

$^5${Graduate School of Information Science, Nara Institute of Science and Technology} 

{\tt ger@is.naist.jp}. 

\end{center}
\bs

\n{\bf Abstract.}\hs{1} 
Since quantum information is continuous, its handling is sometimes surprisingly 
harder than the classical counterpart. A typical example is cloning; making a copy of digital
information is straightforward but it is not possible exactly for
quantum information. The question in this paper is whether or not {\em
quantum} network coding is possible. Its classical counterpart is
another good example to show that digital information flow can be done
much more efficiently than conventional (say, liquid) flow.

Our answer to the question is similar to the case of cloning, namely,
it is shown that quantum network coding is possible if approximation
is allowed, by using a simple network model called Butterfly. In this
network, there are two flow paths, $s_1$ to $t_1$ and $s_2$ to $t_2$,
which shares a single bottleneck channel of capacity one. In the
classical case, we can send two bits simultaneously, one for each
path, in spite of the bottleneck. Our results for quantum network
coding include: (i) We can send any quantum state $|\psi_1\rangle$
from $s_1$ to $t_1$ and $|\psi_2\rangle$ from $s_2$ to $t_2$
simultaneously with a fidelity strictly greater than $1/2$. (ii) If
one of $|\psi_1\rangle$ and $|\psi_2\rangle$ is classical, then the
fidelity can be improved to $2/3$. (iii) Similar improvement is also
possible if $|\psi_1\rangle$ and $|\psi_2\rangle$ are restricted to
only a finite number of (previously known) states. (iv) Several
impossibility results including the general upper bound of the fidelity 
are also given.

\ms

%%\n{\sf Classification of the topic:} Current challenges; quantum computing 

\section{Introduction}
\begin{figure}[tb]\begin{minipage}{0.30\hsize}\begin{center}
\includegraphics*[width=3cm]{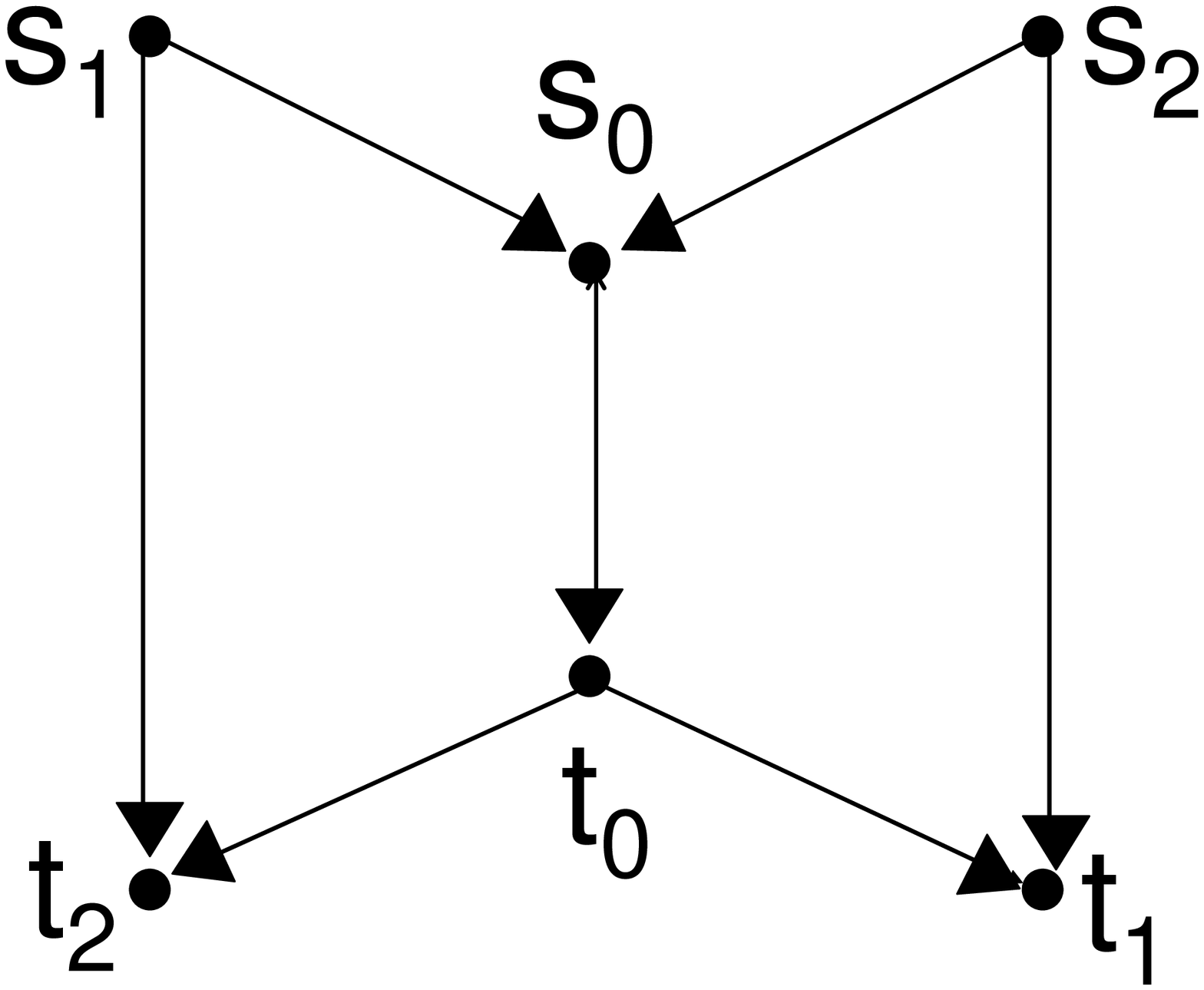}
\caption{Butterfly network.}\label{butterfly} 
\end{center}\end{minipage}
\begin{minipage}{0.30\hsize}
\begin{center}
\includegraphics*[width=3.2cm]{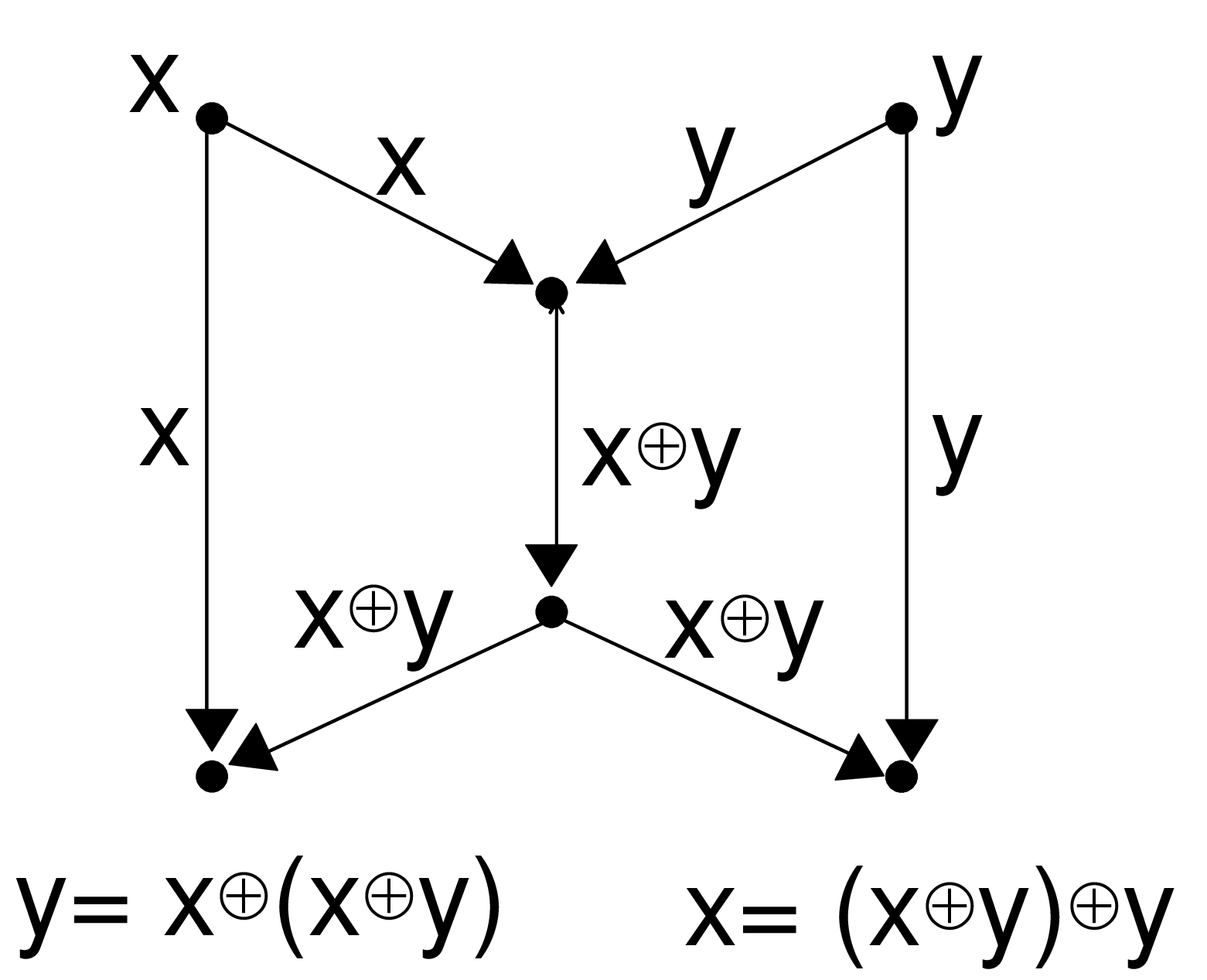}
\caption{Coding scheme}\label{classicalnc} 
\end{center}\end{minipage}
\begin{minipage}{0.30\hsize}
\begin{center}
\includegraphics*[width=3cm]{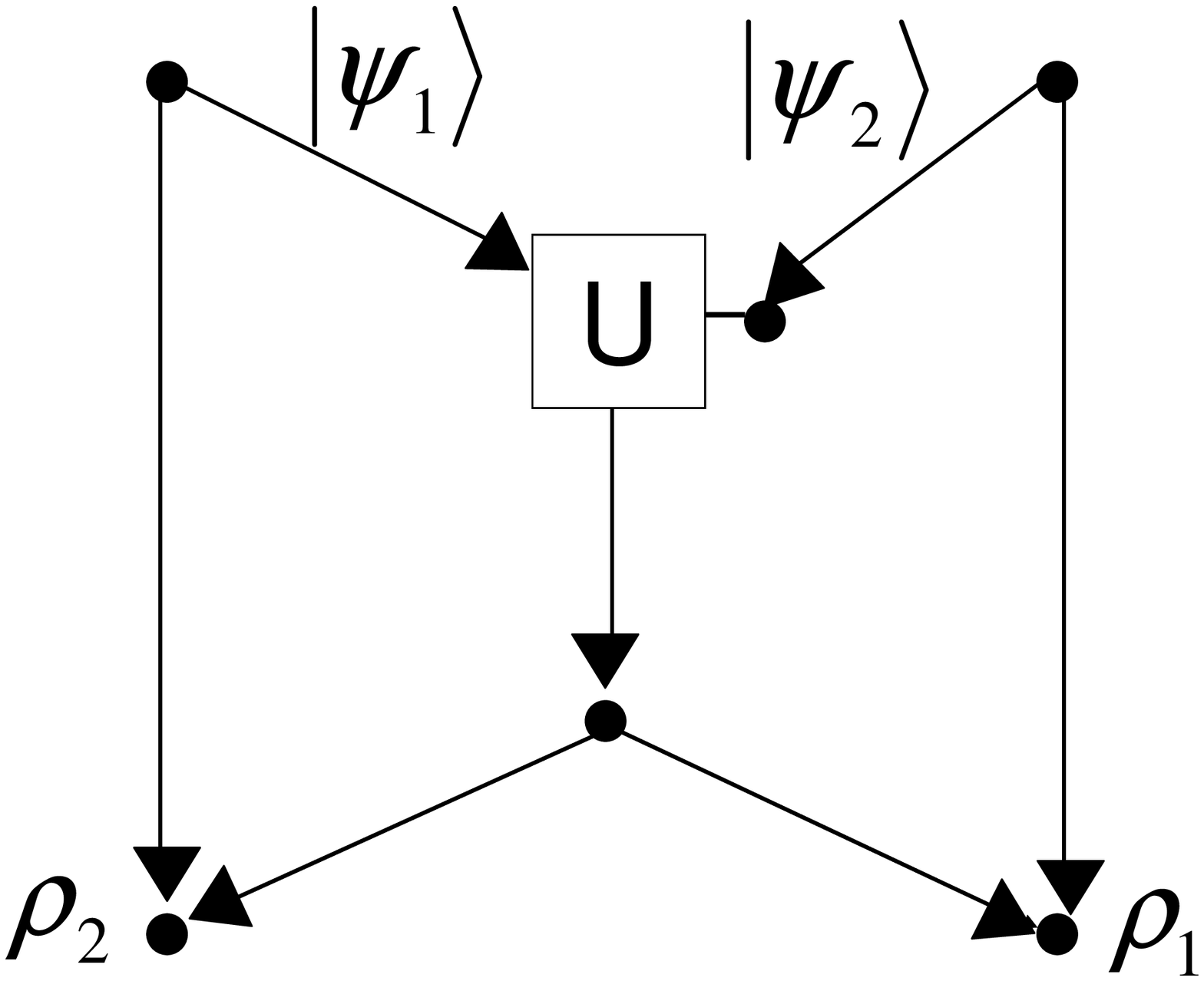}
\caption{Network using a controlled unitary operation}\label{controlU} 
\end{center}\end{minipage}
\end{figure}

In \cite{ACLY00}, Ahlswede, Cai, Li and Yeung showed that the
fundamental law for network flow, the max-flow min-cut theorem, no
longer applies for ``digital information flow.''  The simple and nice
example in \cite{ACLY00} is called the Butterfly network as illustrated
in Fig.\ \ref{butterfly}.  The capacity of each directed link is all
one and there are two source-sink pairs: $s_1$ to $t_1$ and $s_2$ to
$t_2$.  Notice that both paths have to use the single link from $s_0$
to $t_0$ and hence the total amount of (conventional commodity) flow in
both paths is bounded by one, say, $1/2$ for each.  In the case of
digital information flow, however, the protocol shown in
Fig.\ \ref{classicalnc} allows us to transmit two bits, $x$ and $y$,
simultaneously.  Thus, we can effectively achieve larger channel
capacity than what can be achieved by simple routing.  This is known as
{\it network coding} since \cite{ACLY00} and has been quite popular
(see e.g., Network coding home page \cite{NCH} or \cite{AHJKL06,HKM05,Jag05,LL04,LL05} 
for recent developments).

Network coding obviously exploits the two side links, $s_1$ to
$t_2$ and $s_2$ to $t_1$, which are completely useless
graph-topologically. Now the primary question in this paper is whether
this is also possible for {\it quantum} information: Our model is the 
same butterfly network with (unit-capacity) quantum channels and our
goal is to send two qubits from $s_1$ to $t_1$ and $s_2$ to $t_2$
simultaneously. To this end, one should notice that the protocol 
in Fig.~\ref{classicalnc} uses (at least) two tricks. 
One is the EX-OR (Exclusive-OR) operation at node $s_0$;
one can see that the bit $y$ is encoded by using $x$ as a key which is
sent directly from $s_1$ to $t_2$, and vise versa.  The other is the
exact copy of one-bit information at node $t_0$. Are there 
any quantum counterparts for these key operations?

Neither seems easy in the quantum case: For the copy operation, there is the famous no-cloning theorem.  
Also, there is no obvious way of encoding a quantum state by a 
quantum state at $s_0$.  Consider, for example, a simple extension 
of the classical operation at node $s_0$,i.e., a controlled unitary transform $U$ as illustrated 
in Fig.\ \ref{controlU}.  (Note that classical EX-OR is realized by setting
$U=X$ ``bit-flip.'') Then, for any $U$, there is a quantum state
$|\phi\rangle$ (actually an eigenvector of $U$) such that
$|\phi\rangle$ and $U|\phi\rangle$ are identical (up to a global
phase).  Namely, if $|\psi_1\rangle=|\phi\rangle$, then the quantum
state at the output of $U$ is exactly the same for $|\psi_2\rangle=|0\rangle$ 
and $|\psi_2\rangle = |1 \rangle$.  This means their difference is completely lost 
at that position and hence is completely lost at $t_1$ also.

Thus it is highly unlikely that we can achieve an exact transmission
of two quantum states, which forces us to consider an {\it
approximate} transmission.  As an approximation factor, we use a
(worst-case) {\it fidelity} between the input state $|\psi_1\rangle$
at $s_1$ ($|\psi_2\rangle$ at $s_2$, resp.) and the output state
$\pmb{\rho}_1$ at $t_1$ ($\pmb{\rho}_2$ at $t_2$, resp.)  Recall that
the fidelity is at most 1.0 by definition and 0.5 is automatically
achieved by outputting a completely mixed state.  Thus our question is 
whether we can achieve a fidelity of strictly greater than 0.5.

{\bf Our Contribution.} This paper gives a positive answer to this
question.  We first show that we do need the (topologically useless)
side channels for our goal exactly as in the classical case (Theorem \ref{x-net}).  
Namely, without them, we can prove that for any protocol, there exists a quantum state
$|\psi_i\rangle$ ($i=1$ or $2$) and its output state $\pmb{\rho}_i$ 
such that $F(|\psi_i\rangle, \pmb{\rho}_i) \leq 1/2$.  We then give our protocol
which achieves a fidelity of strictly greater than 1/2 for the
butterfly network (Theorem \ref{thm-2qubits}).  The idea is discretization of (continuous) quantum
states. Namely, the quantum state from $s_2$ is changed into classical
two bits by using what we call ``tetra measurement.''  Those two bits
are then used as a key to encode the state from $s_1$ at node $s_0$
(``group operation'') and also to decode it at node $t_1$.  Our
protocol also depends upon the approximate cloning by Bu\v{z}ek and
Hillery \cite{BH96}.  This obviously distorts quantum states, but
interestingly, it also has a merit (creating entanglement between cloned
states) by which we can handle the second problem on the state 
distinguishability previously mentioned.

Note that the present general lower bound for the fidelity 
is only slightly better than $1/2$ (some $0.52$).  However, if we impose restriction, the
value becomes much better.  For example, if $|\psi_1\rangle$ is a
classical state (i.e. either $|0 \rangle$ or $|1 \rangle$), then the
fidelity becomes 2/3 (Theorem \ref{thm-qubitbit}). Similar improvement
is also possible if $|\psi_1\rangle$ and $|\psi_2\rangle$ are
restricted to only a finite number of (previously known) states,
especially if they are the so-called quantum random access coding states
\cite{ANTV02}. By using those states, we can design an interesting protocol 
which can send two classical bits from $s_1$ to $t_1$ (similarly two 
bits from $s_2$ to $t_2$) but only one of them, determined by adversary, should be recovered. 
It is shown that the success probability for this protocol is $1/2+\sqrt{2}/16$ (Theorem
\ref{thm-2bits}), but classically the success probability for any protocol is at most $1/2$.

On the negative side, several upper bounds for the fidelity are
given.  Again, the most general one (Theorem \ref{xqq}) may not seem
very impressive (some 0.983), but it is improved under restrictions.
In particular, if we impose the {\it BC} (bit-copy) assumption, we can prove
an upper bound of $11/12$ (Theorem \ref{natural-bound}). (BC means
that whenever we need to copy a classical bit, we use the classical
(exact) copy, which seems quite reasonable.)  We also give a limit of
transmitting random access coding states.  Note that Theorem
\ref{thm-2bits} can be extended to the three-bit case (with success
probability some 0.525) but that is the limit; no protocol exists for
the four-bit transmission with success probability strictly greater
than $1/2$ (Theorem \ref{4-1code}).

{\bf Related Work.} We usually allow approximation and/or errors in quantum computation,
which seems to be an essence of its power in some occasions. 
One example is observed in communication complexity: The quantum
communication complexity to compute the equality function $EQ_n$
exactly is $n$ \cite{HW02}. However, even one qubit communication
enables us to compute $EQ_n$ with success probability larger than $1/2$.
Another example can be seen in locally decodable codes and private information
retrievals: Any $2n$-bit Boolean function $F$ can be computed
with success probability $>1/2$ from an $(n+1)$-qubit information
\cite{WW05}.  Namely, $n+1$ qubits can encode $2n$ classical bits for
computing {\it any} Boolean function approximately.

Thus ``$1/2 + \epsilon$ for very small $\epsilon$'' seems very powerful.
Interestingly, this is not the case in some other occasions. the Nayak
bound \cite{Nay99} says that there is no way to send two bits by one
qubit with success probability $>1/2$.  Moreover, \cite{HINRY06} shows that
one-qubit random access coding for four bits can only be done with 
success probability at most $1/2$, although we can enjoy a good success
probability up to three bits. In this context, our model in this paper also shows 
a clear difference depending on whether or not the two side links exist.

The study of coding methods on quantum information and computation 
has been deeply explored for error correction of quantum computation 
(since \cite{Sho95}) and data compression of quantum sources (since \cite{Sch95}). 
Recall that their techniques are duplication of data (error correction) 
and average-case analysis (data compression). 
Those standard approaches do not seem to help in the core of our problem. 
More tricky applications of quantum mechanism are quantum
teleportation \cite{BBC+93}, superdense coding \cite{BW92}, and a
variety of quantum cryptosystems including the BB84 key distribution
\cite{BB84}. The random access coding by Ambainis, Nayak, Ta-shma, and Vazirani \cite{ANTV02} 
is probably most related one to this paper, which allows us to encode 
two or more classical bits into one qubit and decode it to recover any one of the source bits. Our third
protocol is a realization of this scheme on the Butterfly network. 

The introduction of quantum network coding \cite{HINRY06-qip} 
triggered several new studies: Leung, Oppenheim, and Winter \cite{LOW06} 
examined the asymptotic relation between the amount of quantum information 
and channel capacities on the Butterfly network (and more). 
Shi and Soljanin \cite{SS06} considered multicasting networks 
from the viewpoint of lossless compression and decompression of copies of quantum states. 

\section{The Model} 
\begin{figure}[tb]
\begin{minipage}{0.5\hsize}\begin{center}
\includegraphics*[width=6cm]{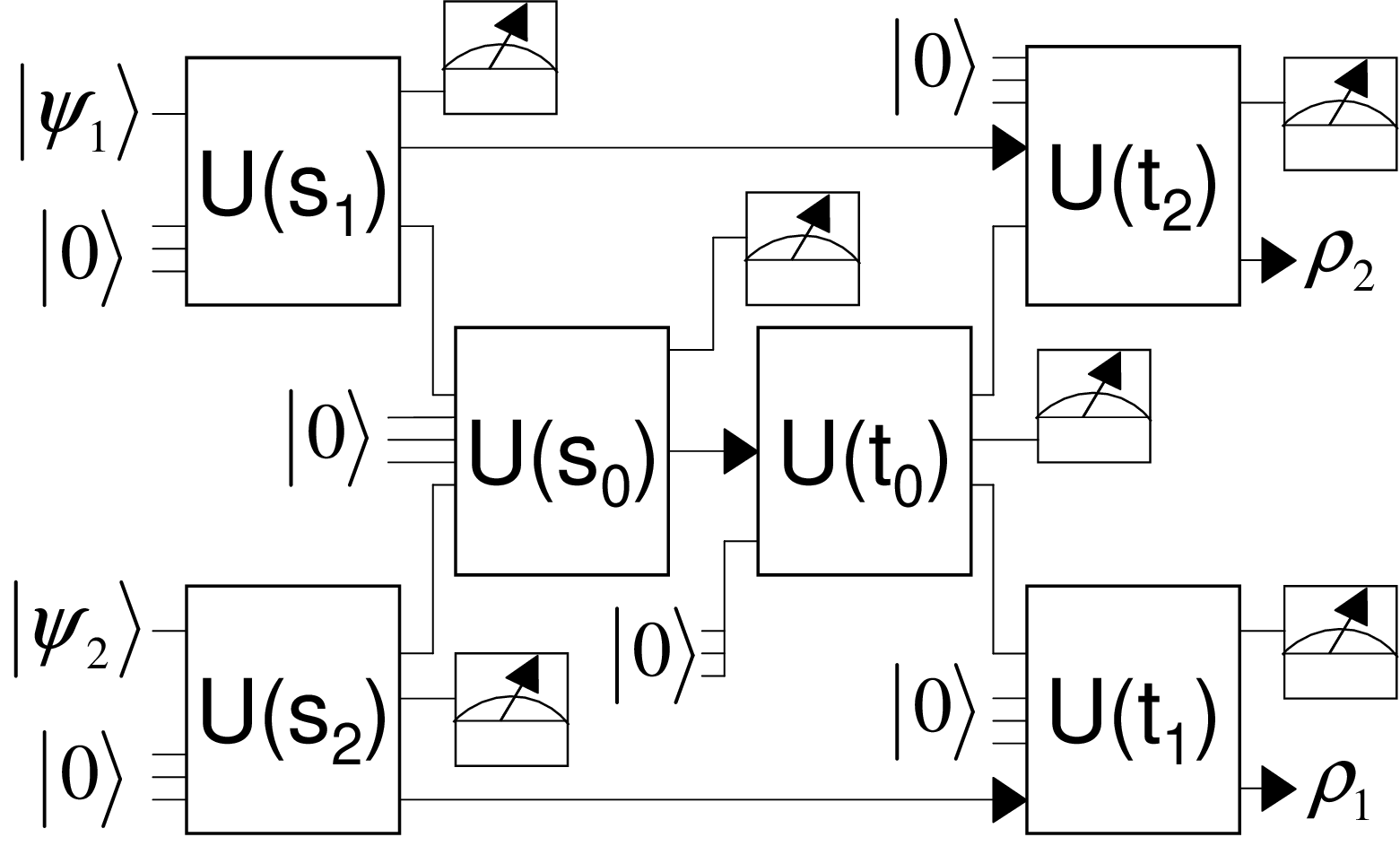}
\caption{Quantum circuit for coding on the Butterfly network}\label{circuit}
\end{center}\end{minipage}
\begin{minipage}{0.5\hsize}
\begin{center}
\includegraphics*[width=3.7cm]{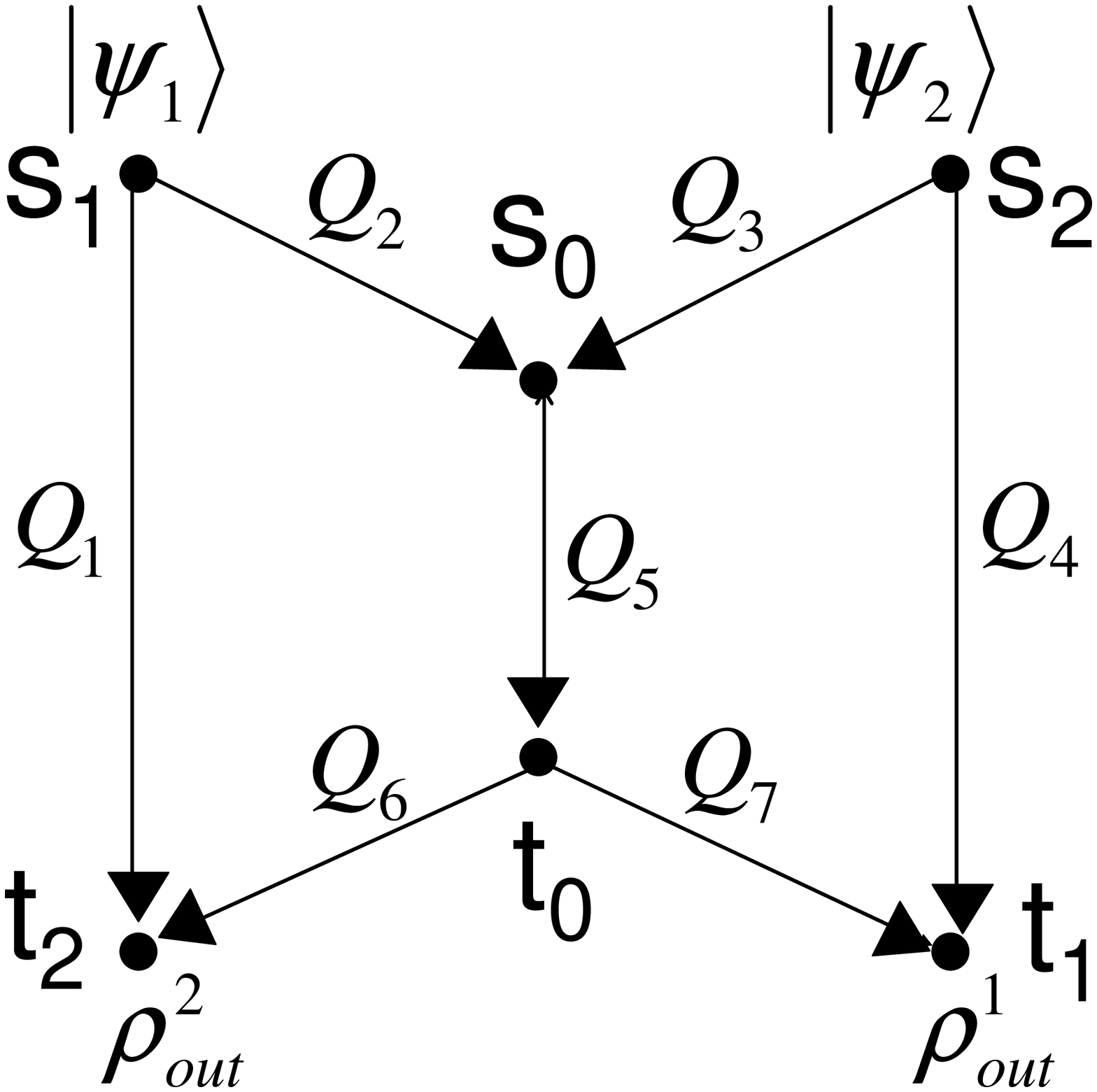}
\caption{Protocol $XQQ$.}\label{twoqubits}
\end{center}\end{minipage}
\end{figure}

Our model as a quantum circuit is shown in Fig.\ \ref{circuit}. 
The information sources at nodes $s_1$ and $s_2$ are
pure one-qubit states $|\psi_1\rangle$ and $|\psi_2\rangle$. (It turns
out, however, that the result does not change for mixed states because
of the joint concavity of the fidelity \cite{NC00}.) Any node does not
have prior entanglement with other nodes. At every node, a physically
allowable operation, i.e., trace-preserving completely positive map
(TP-CP map), is done, and each edge can send only one qubit. They are
implemented by unitary operations with additional ancillae and by
discarding all qubits except for the output qubits \cite{AKN98,NC00}.
   
Our goal is to send $|\psi_1\rangle$ to node $t_1$ and
$|\psi_2\rangle$ to node $t_2$ as well as possible. The quality of
data at node $t_j$ is measured by the fidelity between the original
state $|\psi_j\rangle$ and the state $\pmb{\rho}_j$ output at node
$t_j$ by the protocol. Here, the fidelity between two quantum states
$\pmb{\rho}$ and $\pmb{\sigma}$ are defined as
$F(\pmb{\sigma},\pmb{\rho})=\left(\mathrm{Tr}\sqrt{\pmb{\rho}^{1/2}\pmb{\sigma}\pmb{\rho}^{1/2}}\right)^2$
as in \cite{BHW,Bru00,FMWI02}. (The other common definition is
$\mathrm{Tr}\sqrt{\pmb{\rho}^{1/2}\pmb{\sigma}\pmb{\rho}^{1/2}}$.) In
particular, the fidelity between a pure state $|\psi\rangle$ and a
mixed state $\pmb{\rho}$ is $F(|\psi\rangle,\pmb{\rho})=
\langle\psi|\pmb{\rho}|\psi\rangle$. (To simplify the description, for
a pure state $|\psi\rangle\langle\psi|$ we often use the vector
representation $|\psi\rangle$ and we also use bold fonts for a
$2\times2$ or $4\times4$ density matrix for exposition.) 
We call the minimum of $F(|\psi_1\rangle,\pmb{\rho}_1)$ over all one-qubit states
$|\psi_1\rangle$ and $|\psi_2\rangle$ the {\em fidelity at node $t_1$}
and similarly for {\em fidelity at node $t_2$}.

Before presenting our protocols achieving a fidelity of strictly
greater than $1/2$, we show that the two side links, which are useless
graph-topologically, are indispensable.  One might think this is
trivial from the Nayak bound \cite{Nay99}.  Namely, if the two inputs
are classical $0/1$ bits, then they cannot be sent using a single quantum
channel ($s_0$ to $t_0$) with success probability ($=$ fidelity) greater than
$1/2$.  This is not true since our definition only requires a fidelity
at {\it each} sink.  In fact, we can achieve a fidelity of at least
$0.75$ in our definition, by simply using the one-qubit random access coding 
for two bits \cite{ANTV02} and the phase-covariant cloning (a kind of 
approximated cloning) \cite{Bru00,FMWI02}. (Note that $0.75^2 > 0.5$ but this does not violate 
the Nayak bound since the success probabilities at the two sides are not 
independent.) The proof of the following theorem needs a careful consideration 
of physical operations on the Bloch ball (see, e.g., \cite{FA99,RSW02}) and the trace distance. 
Notice that in this paper, the trace distance between two quantum states $\pmb{\rho}$ and $\pmb{\rho'}$ 
is defined to be $||\pmb{\rho}-\pmb{\rho'}||_{tr}$ without the normalization factor $2$ as in \cite{NC00}. 
(If two states are qubits, this distance is equal to the geometrical 
distance of the corresponding points in the Bloch ball.)

\begin{theorem}\label{x-net}
No quantum protocol can achieve fidelity larger than $1/2$ 
if both side links are removed from the Butterfly. 
\end{theorem}

\begin{proof}
We show that, for any proper protocol, if the fidelity at $t_2$ is larger than $1/2$ 
(say, $1/2+\epsilon$ with $\epsilon>0$) then the fidelity at $t_1$ is strictly less than $1/2$. 
For our purpose, we consider the case where the sources at $s_1$ and $s_2$ are a qubit 
$|\psi\rangle$ and a classical bit $b$, respectively. 
We can assume that they are sent to $s_0$ without any transformation 
(since otherwise their operations at $s_1$ and $s_2$ can be delayed until $s_0$). 
Now, let ${\cal E}_b$ be the images of the Bloch ball 
resulting from operations at $s_0$ when $b$ is sent from $s_2$. 
Let the distance between ${\cal E}_0$ and ${\cal E}_1$ be the minimum trace distance 
between any state in ${\cal E}_0$ and that in ${\cal E}_1$. 
Then, the following lemma holds from the fidelity requirement at $t_2$:
\begin{lemma}\label{image-distance}
The distance between ${\cal E}_0$ and ${\cal E}_1$ is at least $4\epsilon$. 
\end{lemma} 
\begin{proof}
Let $C_b$ be the TP-CP map at $s_0$ when $b$ is sent from $s_2$. 
We can regard the operations at $t_0$ and $t_2$ along the path $s_2$-$t_2$ 
as the measurement defined by a POVM (positive operator-valued measure) $\{E_0,E_1\}$. 
(Recall that any measurement is defined by a POVM $\{E_i\}_i$, that is, 
each operator $E_i$ is positive and $\sum_i E_i =I$.) 
Then, to prove the lemma we need to show that for any one-qubit states 
$|\psi\rangle,|\psi'\rangle$, 
 $||C_0(|\psi\rangle)-C_1(|\psi'\rangle)||_{tr}\geq 4\epsilon$. 
However, by the fidelity requirement at $t_2$, for any 
 $|\psi\rangle,|\psi'\rangle$ and any $b=0,1$, it must hold that 
 $\mathrm{Tr}(E_bC_b(|\psi\rangle))\geq 1/2+\epsilon$  
and $\mathrm{Tr}(E_bC_{1-b}(|\psi'\rangle))\leq 1/2-\epsilon$. 
Thus, we have $||C_0(|\psi\rangle)-C_1(|\psi'\rangle)||_{tr}\geq 
\sum_{b=0,1}|\mathrm{Tr} E_b( C_0(|\psi\rangle)-C_1(|\psi'\rangle) ) |
\geq 4\epsilon$, where the first inequality is obtained from the following fact:  
For any quantum states $\pmb{\rho},\pmb{\rho}'$, $||\pmb{\rho}-\pmb{\rho}'||_{tr}$ 
equals \sloppy $\mathrm{max}_{\{F_0,F_1\}}\sum_b|\mathrm{Tr}F_b(\pmb{\rho}-\pmb{\rho}')|$ 
where the maximization is over all POVMs $\{F_0,F_1\}$ \cite{NC00}. 
\end{proof}
By Lemma \ref{image-distance}, the center of the Bloch ball is outside at least one of ${\cal E}_0$ and ${\cal E}_1$. 
Now let ${\cal F}_b$ be the final images at $t_1$ when $b$ is the source at $s_2$, 
and let $T$ be the composite TP-CP map of $t_0$ and $t_1$ along the path $s_1$-$t_1$. 
Note that ${\cal F}_0=T({\cal E}_0)$ and ${\cal F}_1=T({\cal E}_1)$, and $T$ is linear. 
Since $T$ transforms the Bloch ball into an ellipsoid within the Bloch ball \cite{FA99,RSW02}, 
the center of the Bloch ball is outside one of ellipsoids ${\cal F}_0$ and ${\cal F}_1$, 
say ${\cal F}_0$. This means that there exists some input state $|\psi\rangle$ at $s_1$ such 
that $|\psi\rangle$ and its output state $\pmb{\rho}_\psi$ at $t_1$ are 
in different half of the Bloch ball, that is, $F(|\psi\rangle,\pmb{\rho}_\psi)<1/2$. 
Therefore the fidelity at $t_1$ is $<1/2$.
\end{proof}

\section{Protocol for Crossing Two Qubits}
In this section we prove the following lower bound.

\begin{theorem}\label{thm-2qubits}
There exists a quantum protocol whose fidelities at nodes $t_1$ and $t_2$ 
are $1/2+2/81$ and $1/2+2\sqrt{3}/243$, respectively. 
\end{theorem}
  
\subsection{Overview of the Protocol}
Fig.\ \ref{twoqubits} illustrates our protocol, Protocol for Crossing
Two Qubits ($XQQ$). As expected, the approximated cloning is used at
nodes $s_1$, $s_2$ and $t_0$. At node $s_0$, we first apply the tetra
measurement to the state of one-qubit system ${\cal Q}_3$ and obtain
two classical bits $r_1r_2$. Their different four values suggest which
part of the Bloch sphere the state of ${\cal Q}_3$ sits in. These four
values are then used to choose one of four different operations, the group
operations, to encode the state of ${\cal Q}_2$. These four operations
include identity $I$, bit-flip $X$, phase-flip $Z$, and
bit$+$phase-flip $Y$. At node $t_1$, we apply the reverse operations
of these four operations (actually the same as the original ones) for
the decoding purpose.

At node $t_2$, we recover the two bits $r_1r_2$ (actually the
corresponding quantum state for the output state) by comparing ${\cal
Q}_1$ and ${\cal Q}_6$. This should be possible since ${\cal Q}_2$
$(\approx {\cal Q}_1)$ is encoded into ${\cal Q}_5$ $(\approx {\cal
Q}_6)$ by using $r_1r_2$ as a key but its implementation is not
obvious. It is shown that for this purpose, we can use the Bell
measurement together with the fact that ${\cal Q}_1$ and ${\cal Q}_2$
are partially entangled as a result of cloning at node $s_1$.

{\bf Remark.} It is not hard to average the fidelities at $t_1$ and $t_2$ 
by mixing the encoding state at $t_1$ with the Bell state $(|00\rangle+|11\rangle)/\sqrt{2}$, 
implying $1/2+2(2-\sqrt{3})/27\approx 0.52$ at both sinks. 
  
\subsection{Building Blocks}\label{building-block}

{\bf Universal Cloning (UC).} 
As the first tool of our protocol, we recall the notion of the approximated cloning 
by Bu\v{z}ek and Hillery \cite{BH96}, called the {\em universal cloning}.  
Let $|\Psi^+\rangle=\frac{1}{\sqrt{2}}(|01\rangle+|10\rangle)$. 
Then, it is given by the TP-CP map $UC$ defined by   
\begin{align}\label{BHcopy}
&UC(|0\rangle\langle 0|)= \frac{2}{3}|00\rangle\langle 00|+\frac{1}{3}|\Psi^+\rangle\langle\Psi^+|,\ \ 
UC(|0\rangle\langle 1|)= \frac{\sqrt{2}}{3}|\Psi^+\rangle\langle 11|
+\frac{\sqrt{2}}{3}|00\rangle\langle\Psi^+|,\nonumber\\
&UC(|1\rangle\langle 0|)= \frac{\sqrt{2}}{3}|11\rangle\langle\Psi^+|
+\frac{\sqrt{2}}{3}|\Psi^+\rangle\langle 00|,\ \ 
UC(|1\rangle\langle 1|)= \frac{2}{3}|11\rangle\langle 11|+ \frac{1}{3}|\Psi^+\rangle\langle\Psi^+|.
\end{align}
This map is intended to clone not only classical states $|0\rangle$ and $|1\rangle$ but also any superposition 
equally well by mixing the symmetric state $|\Psi^+\rangle$ with $|00\rangle$ and $|11\rangle$ as the output. 
Let $\pmb{\rho}_1=\mathrm{Tr}_2 UC(|\psi\rangle)$ and $\pmb{\rho}_2=\mathrm{Tr}_1 UC(|\psi\rangle)$, 
where $\mathrm{Tr}_i$ is the partial trace over the $i$-th qubit. 
Then, easy calculation implies that $\pmb{\rho}_1=\pmb{\rho}_2=
\frac{2}{3}|\psi\rangle\langle\psi|+\frac{1}{3}\cdot\frac{\pmb{I}}{2}$, 
which means $F(|\psi\rangle,\pmb{\rho}_1)=F(|\psi\rangle,\pmb{\rho}_2)=5/6$. 
We call its induced map $|\psi\rangle\mapsto\pmb{\rho}_1$ (or $|\psi\rangle\mapsto\pmb{\rho}_2$) 
the {\em universal copy}.  

{\bf Tetra Measurement (TTR).} Next, we introduce the tetra measurement. 
We need the following four states 
$|\chi(00)\rangle=\cos\tilde{\theta}|0\rangle
+e^{\imath\pi/4}\sin\tilde{\theta}|1\rangle$, 
$|\chi(01)\rangle=\cos\tilde{\theta}|0\rangle    
+e^{-3\imath\pi/4}\sin\tilde{\theta}|1\rangle$, 
$|\chi(10)\rangle=\sin\tilde{\theta}|0\rangle    
+e^{-\imath\pi/4}\cos\tilde{\theta}|1\rangle$, and 
$|\chi(11)\rangle=\sin\tilde{\theta}|0\rangle 
+e^{3\imath\pi/4}\cos\tilde{\theta}|1\rangle$ 
with $\cos^2\tilde{\theta}=1/2+\sqrt{3}/6$, which form a tetrahedron in the Bloch sphere representation. 
The {\em tetra measurement}, denoted by $TTR$, is the POVM defined
by \sloppy $\{\frac{1}{2}|\chi(00)\rangle\langle\chi(00)|,
\frac{1}{2}|\chi(01)\rangle\langle\chi(01)|, \frac{1}{2}|\chi(10)\rangle\langle\chi(10)|,
\frac{1}{2}|\chi(11)\rangle\langle\chi(11)|\}$.  

{\bf Group Operation (GR).} In what follows, let $X=\left(\begin{array}{cc}
0 & 1\\1 & 0
\end{array}\right)$ be the bit-flip operation, $Z=\left(
\begin{array}{cc}
1 & 0 \\0 & -1
\end{array}
\right)$ be the phase-flip operation, and $Y=XZ$.  
Notice that the set of unitary maps on one-qubit states $\pmb{\rho}\mapsto W\pmb{\rho}W^\dagger$ $(W=I,Z,X,Y)$ 
is the Klein four group. The {\em group operation under} a two-bit string $r_1r_2$, 
denoted by $GR(\pmb{\rho},r_1r_2)$, is a transformation defined by $GR(\pmb{\rho},00)=\pmb{\rho}$, 
$GR(\pmb{\rho},01)=Z\pmb{\rho}$, $GR(\pmb{\rho},10)=X\pmb{\rho}$, and $GR(\pmb{\rho},11)=Y\pmb{\rho}$. 
Note that we frequently use simplified expressions like $X\pmb{\rho}$ instead of $X\pmb{\rho} X^\dagger$. 

{\bf 3D Bell Measurement (BM).} Moreover, for recovering $|\psi_2\rangle$ at node $t_2$ 
we introduce another new operation based on the Bell measurement, 
$BM({\cal Q},{\cal Q}')$ (or $BM(\pmb{\sigma})$), which applies the following three operations (a), (b), and (c) 
with probability $1/3$ for each, to the state $\pmb{\sigma}$ (a $4\times 4$ density matrix) 
of the two-qubit system ${\cal Q}\otimes {\cal Q}'$.  

(a) \sloppy Measure $\pmb{\sigma}$ in the Bell basis 
$\left\{|\Phi^+\rangle=\frac{|00\rangle+|11\rangle}{\sqrt{2}},|\Phi^-\rangle=\frac{|00\rangle-|11\rangle}{\sqrt{2}},
  |\Psi^+\rangle =\frac{|01\rangle+|10\rangle}{\sqrt{2}}
,|\Psi^-\rangle=\frac{|01\rangle-|10\rangle}{\sqrt{2}}
\right\}$,  
and output $|0\rangle$ if the measurement result for $|\Phi^+\rangle$ or $|\Phi^-\rangle$ is obtained, 
and $|1\rangle$ otherwise. 

(b) Measure $\pmb{\sigma}$ similarly, and 
output $|+\rangle$ if the measurement result for $|\Phi^+\rangle$ or $|\Psi^+\rangle$ is obtained, 
and $|-\rangle$ otherwise. 

(c) Measure $\pmb{\sigma}$ similarly, and output $|+'\rangle=\frac{1}{\sqrt{2}}(|0\rangle+\imath|1\rangle)$ 
if the measurement result for $|\Phi^+\rangle$ or $|\Psi^-\rangle$ is obtained, 
and $|-'\rangle=\frac{1}{\sqrt{2}}(|0\rangle-\imath|1\rangle)$ otherwise. 

\subsection{Protocol $XQQ$ and Its Performance Analysis}\label{xqq-analysis}     
Now here is the formal description of our protocol.

\

\noindent
{\bf Protocol $XQQ$}: Input $|\psi_1\rangle$ at $s_1$, and $|\psi_2\rangle$ at $s_2$; 
Output $\pmb{\rho}_{out}^1$ at $t_1$, and $\pmb{\rho}_{out}^2$ at $t_2$.

Step 1. $({\cal Q}_1,{\cal Q}_2)=UC(|\psi_1\rangle)$ at $s_1$, 
and $({\cal Q}_3,{\cal Q}_4)=UC(|\psi_2\rangle)$ at $s_2$. 

Step 2. ${\cal Q}_5=GR({\cal Q}_2,TTR({\cal Q}_3))$ at $s_0$.  

Step 3. $({\cal Q}_6,{\cal Q}_7)=UC({\cal Q}_5)$ at $t_0$. 

Step 4 (Decoding at node $t_1$ and $t_2$). 
$\pmb{\rho}_{out}^1=GR({\cal Q}_7,TTR({\cal Q}_4))$, 
and $\pmb{\rho}_{out}^2=BM({\cal Q}_1,{\cal Q}_6)$.

\

We give the proof of Theorem \ref{thm-2qubits} by analyzing protocol $XQQ$. 
For this purpose, we introduce the notion of shrinking maps (also known as 
a depolarizing channel \cite{NC00}), which plays an important role 
in the following analysis of $XQQ$: Let $\pmb{\rho}$ be any quantum state. 
Then, if a map $C$ transforms $\pmb{\rho}$ to $p\cdot \pmb{\rho}+(1-p)\frac{\pmb{I}}{2}$ 
for some $0\leq p\leq 1$, then $C$ is said to be {\em $p$-shrinking}. 
The following three lemmas are immediate: 

\begin{lemma}\label{shrink-composition}
If $C$ is $p$-shrinking and $C'$ is $p'$-shrinking, then $C\circ C'$ is $pp'$-shrinking. 
\end{lemma}

\begin{lemma}\label{shrink-fidelity}
If $C$ is $p$-shrinking, $F(\pmb{\rho},C(\pmb{\rho}))\geq 1/2+p/2$ for any state $\pmb{\rho}$. 
\end{lemma}

\begin{lemma}\label{shrink}
The universal copy is $2/3$-shrinking. 
\end{lemma}

{\bf Computing the Fidelity at Node {\boldmath $t_1$}.} We first investigate the quality of the path 
from $s_1$ to $t_1$. Fix $\pmb{\rho}_2=|\psi_2\rangle\langle\psi_2|$ as an arbitrary state at node $s_2$ 
and consider four maps 
$C_1$: $|\psi_1\rangle\rightarrow {\cal Q}_2$, 
$C_2[\pmb{\rho}_2]$: ${\cal Q}_2\rightarrow {\cal Q}_5$, 
$C_3$: ${\cal Q}_5\rightarrow {\cal Q}_7$ and 
$C_4[\pmb{\rho}_2]$: ${\cal Q}_7\rightarrow \pmb{\rho}_{out}^1$. 
We wish to compute the composite map $C_{s_1t_1}=C_4[\pmb{\rho}_2]\circ C_3
\circ C_2[\pmb{\rho}_2]\circ C_1$ and its fidelity. 
We need two more lemmas before the final one (Lemma \ref{final-1}). 

\begin{lemma}\label{c3c2}
$C_3\circ C_2[\pmb{\rho}_2]=C_2[\pmb{\rho}_2]\circ C_3$. 
\end{lemma}

\begin{proof}%{Lemma \ref{c3c2}}
We decompose the $p$-shrinking map $C^{(p)}$ into $C^{(p)}=pC^{(1)}+(1-p)C^{(0)}$. 
Here, the $1$-shrinking map $C^{(1)}$ is the identity and 
the $0$-shrinking map $C^{(0)}$ transforms any state to $\frac{\pmb{I}}{2}$. 
Clearly, $C^{(1)}$ is commutative with $C_2[\pmb{\rho}_2]$. 
For showing the commutativity between $C^{(0)}$ and $C_2[\pmb{\rho}_2]$, we prove 
that, for any qubit $\pmb{\rho}$ and any $2\times 2$ matrix basis element 
$|b\rangle\langle b'|$ (where $b,b'\in\{0,1\}$) on ${\cal Q}_3$, 
$C^{(0)}(GR(\pmb{\rho},TTR(|b\rangle\langle b'|)))=GR(C^{(0)}(\pmb{\rho}),TTR(|b\rangle\langle b'|))$. 
The left-hand side is $\sum_{r\in\{0,1\}^2} 
\langle r|TTR(|b\rangle\langle b'|)|r\rangle \frac{\pmb{I}}{2}$ since 
$GR(\pmb{\rho},TTR(|b\rangle\langle b'|))=\sum_{r\in\{0,1\}^2} 
\langle r|TTR(|b\rangle\langle b'|)|r\rangle GR(r,\pmb{\rho})$ 
and $C^{(0)}$ is $0$-shrinking. The right-hand side is 
$$
GR(\pmb{I}/2,TTR(|b\rangle\langle b'|))
=\sum_{r\in\{0,1\}^2}\langle r|TTR(|b\rangle\langle b'|)|r\rangle 
GR(r,\pmb{I}/2) =\sum_{r\in\{0,1\}^2}\langle r|TTR(|b\rangle\langle b'|)|r\rangle \frac{\pmb{I}}{2}.
$$ 
Thus, by linearity, the $p$-shrinking map and $C_2[\pmb{\rho}_2]$ are commutative. 
Now the proof is completed since $C_3$ is $2/3$-shrinking by Lemma \ref{shrink}. 
\end{proof}

\begin{lemma}\label{mainlemma} (Main Lemma) 
$C_4[\pmb{\rho}_2]\circ C_2[\pmb{\rho}_2]$ is $\frac{1}{9}$-shrinking. (See below for the proof.)
\end{lemma} 

\begin{lemma}\label{final-1}
For any $|\psi_1\rangle$, $F(|\psi_1\rangle,C_{s_1t_1}(|\psi_1\rangle))\geq 1/2+2/81$.
\end{lemma}

\begin{proof}
By Lemma \ref{c3c2}, $C_{s_1t_1}=C_4[\pmb{\rho}_2]\circ C_2[\pmb{\rho}_2]\circ C_3\circ C_1$. 
$C_3$ and $C_1$ are both $2/3$-shrinking by Lemma \ref{shrink} and 
$C_4[\pmb{\rho}_2]\circ C_2[\pmb{\rho}_2]$ is $\frac{1}{9}$-shrinking by Lemma \ref{mainlemma}. 
It then follows that $C_{s_1t_1}$ is $\frac{4}{81}$-shrinking by Lemma \ref{shrink-composition} 
and its fidelity is at least $1/2+2/81$ by Lemma \ref{shrink-fidelity}.
\end{proof}\vspace{-8mm}

\begin{proofof}{Lemma \ref{mainlemma}} See Fig.\ \ref{twoqubits} again. 
Since we are discussing $C_4[\pmb{\rho}_2]\circ C_2[\pmb{\rho}_2]$, let 
$\pmb{\rho}_1=\left( 
\begin{array}{ll}
a & b\\c & d
\end{array}
\right)$ be the state on ${\cal Q}_2$, $\pmb{\rho}_2=|\psi_2\rangle\langle\psi_2|=\left(
\begin{array}{ll}
e & f\\g & h
\end{array}
\right)$ be the state at $s_2$ and assume that ${\cal Q}_5={\cal Q}_7$. 
We calculate the state on ${\cal Q}_2\otimes{\cal Q}_3\otimes{\cal Q}_4$, 
the state on ${\cal Q}_5\otimes{\cal Q}_4$ $(={\cal Q}_7\otimes{\cal Q}_4)$ 
and $\pmb{\rho}_{out}^1$ in this order. For ${\cal Q}_2\otimes{\cal Q}_3\otimes{\cal Q}_4$, 
recall that $\pmb{\rho}_2$ is cloned into ${\cal Q}_3$ and ${\cal Q}_4$ and so, 
by Eq.(\ref{BHcopy}) in Sec.\ \ref{building-block}, the state on ${\cal Q}_2\otimes{\cal Q}_3\otimes{\cal Q}_4$ 
is written as
\begin{align}\label{q234}
&\pmb{\rho}_1 \otimes \left[
 \frac{2e}{3}|00\rangle\langle 00|+\frac{e}{6}(|01\rangle+|10\rangle)
(\langle 01|+\langle 10|)\right. + \frac{f}{3}\left((|01\rangle+|10\rangle )\langle 11|+|00\rangle (\langle 01|+\langle 10|) \right) \nonumber\\
&\ \ \ +  \frac{g}{3}\left( |11\rangle (\langle 01|+\langle 10|)+ (|01\rangle+|10\rangle )\langle 00| \right)
 +  \left.\frac{2h}{3}|11\rangle\langle 11|+\frac{h}{6}(|01\rangle+|10\rangle)
(\langle 01|+\langle 10|)\right]\nonumber\\
&=\pmb{\rho}_1\otimes|0\rangle\langle 0|\otimes
\left(\frac{2e}{3}|0\rangle\langle 0|+\frac{f}{3}|0\rangle\langle 1|+\frac{g}{3}|1\rangle\langle 0|
+\frac{1}{6}|1\rangle\langle 1|\right)+  \pmb{\rho}_1\otimes|0\rangle\langle 1|\otimes
\left(\frac{1}{6}|1\rangle\langle 0|+\frac{f}{3}\pmb{I}\right) \nonumber\\
&\ \ \ +\pmb{\rho}_1\otimes|1\rangle\langle 0|\otimes
\left(\frac{1}{6}|0\rangle\langle 1|+\frac{g}{3}\pmb{I}\right)+\pmb{\rho}_1\otimes|1\rangle\langle 1|\otimes
\left(\frac{1}{6}|0\rangle\langle 0|+\frac{f}{3}|0\rangle\langle 1|+\frac{g}{3}|1\rangle\langle 0|
+\frac{2h}{3}|1\rangle\langle 1|\right).
\end{align} 

Then, we apply the group operation to the first two bits of ${\cal Q}_2\otimes{\cal Q}_3
\otimes{\cal Q}_4$. In general, for ${\cal Q}\otimes{\cal Q}'$, 
$GR({\cal Q},TTR({\cal Q}'))$ is given as follows (see Appendix for the proof). 

\begin{lemma}\label{cond} Let $\pmb{\rho}$ be the state on ${\cal Q}$. 
Then, $GR({\cal Q},TTR({\cal Q}'))$ is the following TP-CP map: 
\begin{align*}
&\pmb{\rho}\otimes|0\rangle\langle 0|\mapsto\frac{1}{\sqrt{3}}V(I,Z)\pmb{\rho}
 +\left(1-\frac{1}{\sqrt{3}}\right)\cdot\frac{\pmb{I}}{2},\ \ 
\pmb{\rho}\otimes|1\rangle\langle 1|\mapsto
\frac{1}{\sqrt{3}}V(X,Y)\pmb{\rho}+\left(1-\frac{1}{\sqrt{3}}\right)\cdot\frac{\pmb{I}}{2},\\ 
&\pmb{\rho}\otimes|0\rangle\langle 1|\mapsto
\frac{1}{2\sqrt{3}}(V(I,X)\pmb{\rho}-V(Y,Z)\pmb{\rho}+\imath(V(I,Y)\pmb{\rho}-V(Z,X)\pmb{\rho}) ),\\
&\pmb{\rho}\otimes|1\rangle\langle 0|\mapsto
\frac{1}{2\sqrt{3}}(V(I,X)\pmb{\rho}-V(Y,Z)\pmb{\rho}-\imath(V(I,Y)\pmb{\rho}-V(Z,X)\pmb{\rho}) ). 
\end{align*}
Here, $V(I,Z)\pmb{\rho}=\frac{1}{2}(I\pmb{\rho}+Z\pmb{\rho})$, and $V(X,Y)\pmb{\rho}$, 
$V(I,X)\pmb{\rho}$, $V(Y,Z)\pmb{\rho}$, $V(I,Y)\pmb{\rho}$, and $V(Z,X)\pmb{\rho}$ are similarly defined. 
Those six operations are $\pmb{I}$-invariant (meaning it maps $\pmb{I}$ to itself) TP-CP maps. 
\end{lemma}   

Now the state on ${\cal Q}_5\otimes {\cal Q}_4$ is obtained by applying Lemma \ref{cond} to Eq.(\ref{q234}). 
From now on, we omit the term for $\frac{\pmb{I}}{2}$. 
Namely, if the one-qubit state is $\pmb{\rho}+\alpha\frac{\pmb{I}}{2}$, we only describe $\pmb{\rho}$. 
This is not harmful since any operation in this section is $\pmb{I}$-invariant 
and hence the $\frac{\pmb{I}}{2}$ term can be recovered 
at the end by using the trace property. Thus, the state on ${\cal Q}_5\otimes{\cal Q}_4$ 
looks like \begin{align}\label{formula}
&\frac{1}{\sqrt{3}}V(I,Z)\pmb{\rho}_1\otimes
\left(\frac{2e}{3}|0\rangle\langle 0|+\frac{1}{6}|1\rangle\langle 1|\right)
+\frac{1}{\sqrt{3}}V(I,Z)\pmb{\rho}_1\otimes
\left(\frac{f}{3}|0\rangle\langle 1|+\frac{g}{3}|1\rangle\langle 0| \right)\nonumber\\
&+  
\frac{1}{2\sqrt{3}}V(I,X;I,Y;+)\pmb{\rho}_1\otimes\frac{1}{6}|1\rangle\langle 0|
+\frac{1}{2\sqrt{3}}V(I,X;I,Y;+)\otimes\frac{f}{3}\pmb{I}\nonumber\\
&+
\frac{1}{2\sqrt{3}}V(I,X;I,Y;-)\pmb{\rho}_1\otimes\frac{1}{6}|0\rangle\langle 1|
+\frac{1}{2\sqrt{3}}V(I,X;I,Y;-)\otimes\frac{g}{3}\pmb{I}\nonumber\\
&
+\frac{1}{\sqrt{3}}V(X,Y)\pmb{\rho}_1\otimes
\left(\frac{1}{6}|0\rangle\langle 0|+\frac{2h}{3}|1\rangle\langle 1|\right)
+\frac{1}{\sqrt{3}}V(X,Y)\pmb{\rho}_1\otimes
\left(\frac{f}{3}|0\rangle\langle 1|+\frac{g}{3}|1\rangle\langle 0| \right),
\end{align}
where $V(I,X;I,Y;\pm)\pmb{\rho}=V(I,X)\pmb{\rho}-V(Y,Z)\pmb{\rho} \pm \imath(V(I,Y)\pmb{\rho}-V(Z,X)\pmb{\rho})$,  
%and $V(I,X;I,Y;-)=V(I,X)-V(Y,Z)-\imath(V(I,Y)-V(Z,X))$, 
and the terms such that the state of ${\cal Q}_5$ is $\frac{\pmb{I}}{2}$ are omitted. 

We next transform the state of ${\cal Q}_5\otimes{\cal Q}_4$ to $\pmb{\rho}_{out}^1$ 
by using Lemma \ref{cond} again. For example, $V(I,Z)\pmb{\rho}_1\otimes|0\rangle\langle 0|$ 
is transformed to $\frac{1}{\sqrt{3}}V(I,Z)V(I,Z)\pmb{\rho}_1$. 
To simplify the resulting formula, 
the following lemma is used (see Appendix for its proof). 

\begin{lemma}\label{simplify}
1) $V(I,Z)V(I,Z)\pmb{\rho}_1=V(X,Y)V(X,Y)\pmb{\rho}_1=
\left(
\begin{array}{ll}
a & 0\\0 & d
\end{array}
\right)$.

2) $V(I,Z)V(X,Y)\pmb{\rho}_1=V(X,Y)V(I,Z) \pmb{\rho}_1=
\left(
\begin{array}{ll}
d & 0\\0 & a
\end{array}
\right)$.

3) $V(I,X)V(I,X)\pmb{\rho}_1=V(Y,Z)V(Y,Z)\pmb{\rho}_1=
%\frac{1}{2}
\left(
\begin{array}{ll}
%1 & b+c\\
%b+c & 1
\frac{1}{2} & \frac{b+c}{2}\\
\frac{b+c}{2} & \frac{1}{2}
\end{array}
\right)$.

4) $V(I,X)V(Y,Z)\pmb{\rho}_1=V(Y,Z)V(I,X)\pmb{\rho}_1=
\left(
\begin{array}{ll}
\frac{1}{2} & -\frac{b+c}{2}\\
-\frac{b+c}{2} & \frac{1}{2}
\end{array}
\right)$.

5) $V(I,Y)V(I,Y)\pmb{\rho}_1=V(Z,X)V(Z,X)\pmb{\rho}_1=
\left(
\begin{array}{ll}
\frac{1}{2} & \frac{b-c}{2}\\
\frac{c-b}{2} & \frac{1}{2}
\end{array}
\right)$.

6) $V(I,Y)V(Z,X)\pmb{\rho}_1=V(Z,X)V(I,Y)\pmb{\rho}_1=\left(
\begin{array}{ll}
\frac{1}{2} & \frac{c-b}{2}\\
\frac{b-c}{2} & \frac{1}{2}
\end{array}
\right)$.

7) \sloppy For any two operators $V,V'$ taken from any different two sets of  
$\{V(I,Z),V(X,Y)\}$, $\{V(I,X),V(Y,Z)\}$, and $\{V(I,Y),V(Z,X)\}$, $VV'\pmb{\rho}_1=\frac{\pmb{I}}{2}$. 
\end{lemma} 

Now it is a routine calculation (see Appendix for its sequence) 
to obtain $\pmb{\rho}_{out}^1=\left(
\begin{array}{ll}
m_1 & m_2 \\m_3 & m_4 
\end{array}
\right)$ where $m_1$ through $m_4$ are equations using $a,b,c$ and $d$ 
($e,f,g$ and $h$ disappear). Using the fact that $a+d=1$, 
we have $\pmb{\rho}_{out}^1=\frac{1}{9}\pmb{\rho}_1+\frac{1}{9}\pmb{I}$.
Recovering the completely mixed state omitted in our analysis, 
we obtain $C_4[\pmb{\rho}_2]\circ C_2[\pmb{\rho}_2](\pmb{\rho}_1)=
\frac{1}{9}\pmb{\rho}_1+\frac{8}{9}\cdot\frac{\pmb{I}}{2}$. Thus, the map is $\frac{1}{9}$-shrinking. 
\end{proofof}

{\bf Computing the Fidelity at Node {\boldmath $t_2$}.}
By analyzing the quality of the path from $s_2$ to $t_2$, we have 
$F(|\psi_2\rangle,\pmb{\rho}_{out}^2) \geq 1/2+2\sqrt{3}/243$.
Its analysis is different from the previous one
by the antisymmetry of the protocol. Details are given in Appendix.  

\subsection{Upper Bounds}
The following theorem shows a general upper bound for the fidelity of crossing 
two qubits over Butterfly. Recall that we showed in Sec.\ 1 (also Fig.\ \ref{controlU}) 
that the operation at $s_0$ must not resemble a controlled unitary operation. 
Thus, it must be a more general TP-CP map (unitary operation 
with some ancillae). The basic idea of the proof is by showing that a 
good TP-CP map, the one which results in the protocol with fidelity 
close to $1.0$, can be ``approximated'' by a controlled unitary operation. 
Hence, the fidelity of sending two qubits over Butterfly 
must be bounded away from $1.0$. Similar to the proof of Theorem \ref{x-net}, 
we use a geometric view of the TP-CP map on the Bloch ball. However, it is much 
complicated since we have to consider the side links. See Appendix for details 
(whose technique is similar to Theorem \ref{natural-bound} of the next section 
which the reader is recommended to read first.). 

\begin{theorem}\label{xqq}
Let $q$ be the fidelity of a protocol for crossing two qubits simultaneously. 
Then, $q < 0.983$. 
\end{theorem}

\section{Protocol for Crossing a Qubit and a Bit}
\begin{figure}%[bt]
\begin{minipage}{0.5\hsize}
\begin{center}
\includegraphics*[width=5cm]{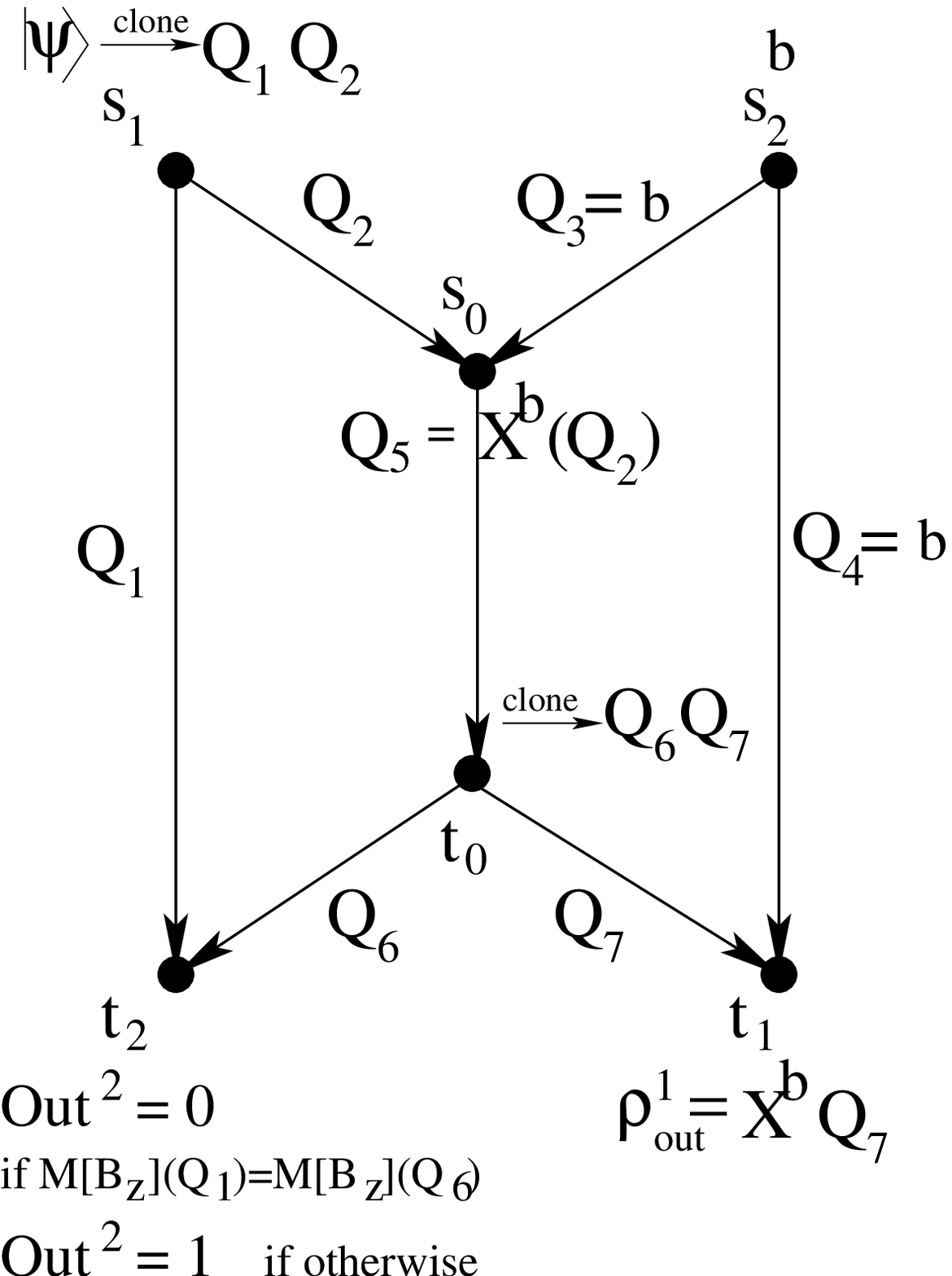}
\caption{Protocol $XQC$}\label{prot1}
\end{center}
\end{minipage}
\begin{minipage}{0.5\hsize}
\begin{center}
\includegraphics*[width=7cm]{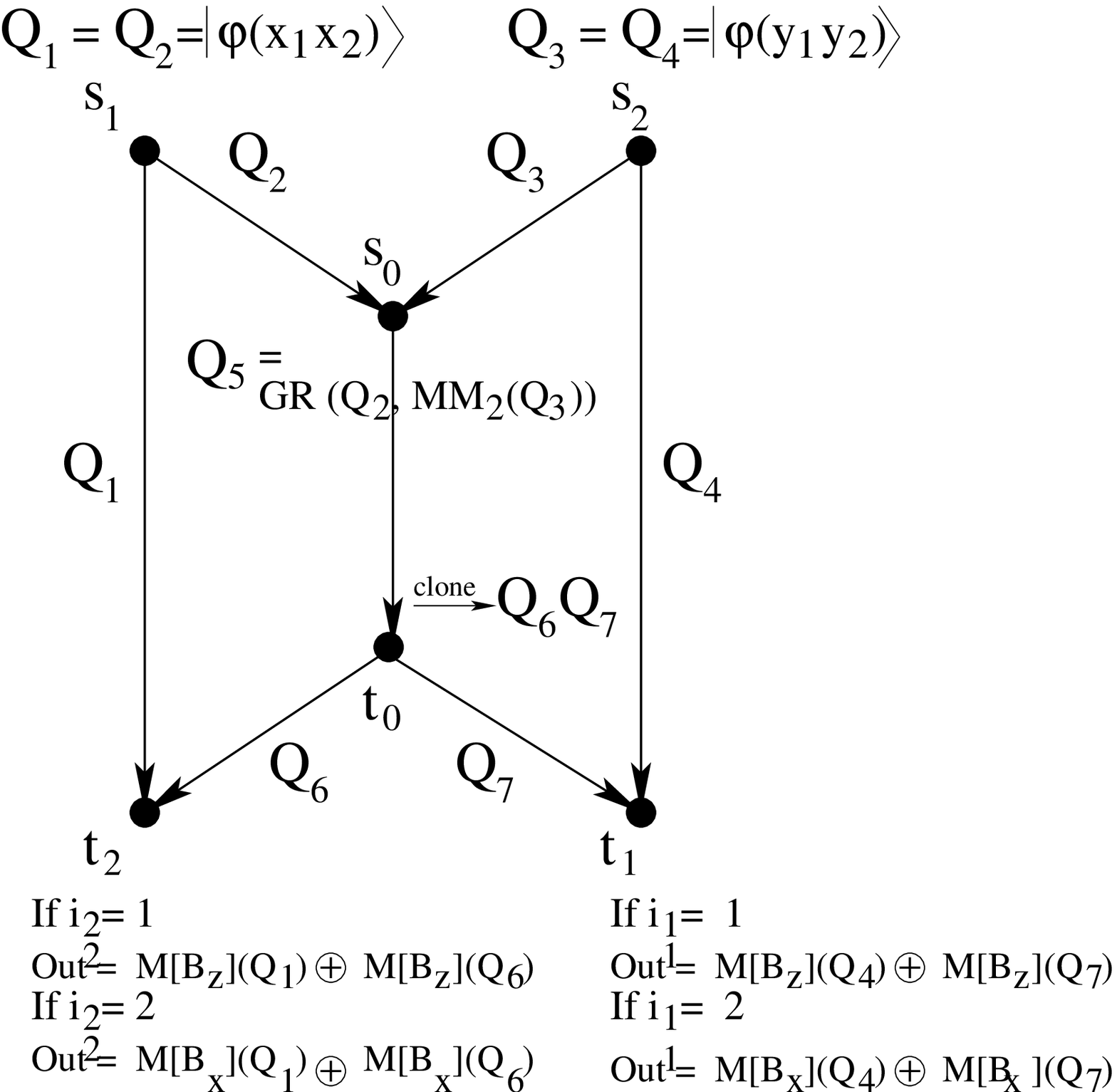}
\caption{Protocol $X2C2C$}\label{prot2}
\end{center}
\end{minipage}
\end{figure}

In this section, we consider the case where one of two sources (say, at $s_2$) 
is a classical bit. Under this situation, we can design a protocol, 
called as $XQC$ (crossing a quantum bit and a classical bit), whose 
fidelity is much better than $XQQ$ (see Appendix for details). The 
protocol is summarized in Fig.~\ref{prot1}, where $M[B_z]({\cal Q})$ means that 
${\cal Q}$ is measured in the basis $B_z=\{|0\rangle,|1\rangle\}$. 
(Similar notations are also used for bases $B_x=\{|+\rangle,|-\rangle\}$ 
and $B_y=\{|+'\rangle,|-'\rangle\}$ later.)  

\begin{theorem}\label{thm-qubitbit} 
$XQC$ achieves the fidelities of $13/18$ and $11/18$ at $t_1$ and
$t_2$. (By averaging the fidelities at both sinks as before, 
we can also have a protocol whose fidelities are the same $2/3$ at 
 $t_1$ and $t_2$.) 
\end{theorem}

On the contrary, assuming that the copies of the bit at $s_2$ are sent to $s_0$ and $t_1$, 
we can obtain an upper bound that is significantly better than Theorem \ref{xqq}. 
In general, this assumption, denoted as the {\em BC (bit-copy) assumption}, is 
reasonable since whenever we need to send a bit to multiple nodes in the 
network, simply sending its (classical) copies does not appear to cause disadvantages.

\begin{theorem}\label{natural-bound}
Let $p$ be the fidelity of a protocol for crossing a bit and a qubit 
under the BC assumption. Then, $p < 11/12$.  
\end{theorem}

\ignore{
\subsection{Proof of Theorem \ref{natural-bound}}
We give the proof of Theorem \ref{natural-bound} before that of Theorem \ref{xqq} 
since both proofs are based on the same idea and the former is simpler 
than the latter.  
}

\begin{proof}%{Theorem \ref{natural-bound}} 
Suppose that there is a protocol whose fidelity is $1-\epsilon$, 
and we wish to show $\epsilon > 1/12$. Here, we give the desired upper bound 
for the case that the capacity of the link from $s_1$ to $t_2$ is unlimited. 
In this case we can assume that the state sent from $s_1$ is pure. 
Let $|\psi\rangle$ and $b$ be the inputs at nodes $s_1$ and $s_2$, respectively. 
By the Schmidt decomposition (see \cite{NC00}), the state after the operation at $s_1$ is written as 
$|\xi\rangle=\alpha|\psi_2\rangle|\psi_1\rangle + \beta|\psi_2^\bot\rangle|\psi_1^\bot\rangle$ 
where $|\psi_1\rangle$ and $|\psi_1^\bot\rangle$ are single-qubit orthonormal states 
on the link to $s_0$ and $|\psi_2\rangle$ and $|\psi_2^\bot\rangle$ are the remaining 
(possibly multi-qubit) orthonormal states on the link to $t_2$. 
Note that $\alpha$, $\beta$, $|\psi_2\rangle$ and $|\psi_1\rangle$ depend on the input $|\psi\rangle$ at $s_1$. 
Without loss of generality, we assume $|\alpha|\geq |\beta|$ (and hence $|\beta|^2\leq 1/2$). 

We first investigate the fidelity on the path from $s_1$ to $t_1$, which is done 
by the following sequence of definitions and observations: 
(i) By the above definition of $|\xi\rangle$, we can write the state on ${\cal Q}_2$ 
(where we use the notations in Fig.\ \ref{prot1} again) as 
$\pmb{\rho}=|\alpha|^2 |\psi_1\rangle\langle\psi_1| + |\beta|^2 |\psi_1^\bot\rangle\langle\psi_1^\bot|$. 
(ii) Intuitively, the value of $|\beta|$ shows the strength of entanglement between ${\cal Q}_1$ 
and ${\cal Q}_2$; if it is large then the distortion of $\pmb{\rho}$ compared to the original 
$|\xi\rangle$ must also be large. In other words, $\beta$ must be small to obtain a small $\epsilon$. 
(iii) For $b=0$ and $1$, let $C_b:$ ${\cal Q}_2\rightarrow {\cal Q}_5$ be the TP-CP map at $s_0$. 
Let $C'_b$ be the corresponding affine map on Bloch-sphere states. Namely, $C'_b$ maps a Bloch vector 
$\vec{r}$ to $O_b^1\Lambda_b O_b^2\vec{r}+\vec{d}_b$, where $O_b^1$ and $O_b^2$ are orthogonal matrices 
with determinant 1, and $\Lambda_b$ is a diagonal matrix (see \cite{FA99,RSW02}). 
(iv) Let $U'_b$ be the map that transforms $\vec{r}$ to $O_b^1O_b^2\vec{r}$. Then, 
we can define the map $U_b$ such that its Bloch-sphere correspondence is $U'_b$. 
Note that $U_b$ is unitary. (v) Let $k_b$ be the distance between the images 
of the Bloch ball by $C'_b$ and $U'_b$. Note that $||(C_b-U_b)|\phi\rangle\langle\phi| ||_{tr}\leq k_b$ 
for an arbitrary pure state $|\phi\rangle$ (where the trace norm $||\cdot||_{tr}$ is defined by $||A||_{tr}=\sqrt{AA^\dagger}$). By a similar reason as (ii) $k_b$ must be small 
for a small $\epsilon$. (vi) Now we select the state $\pmb{\rho}$ which is undesirable 
to achieve a high fidelity, i.e., the one such that $U_0\pmb{\rho}=U_1\pmb{\rho}$ 
(such $\pmb{\rho}$ exists, which is parallel to the eigenvector of $U_0^{-1}U_1$). 
Let $\pmb{\rho'}=|\alpha|^2|\psi_1\rangle\langle\psi_1|$, which is an approximation 
of $\pmb{\rho}$ represented as a product state. (vii) The operation at $t_0$ is considered 
to be the two TP-CP maps on the one qubits: One map $CP_1$ is for $t_1$ and 
the other $CP_2$ is for $t_2$. Their Bloch-sphere correspondence $CP'_1$ and $CP'_2$ 
have a trade-off on the size of their images. So, the image of $CP'_1$ must be large 
for a small $\epsilon$, and then we have a shrinking factor for $CP'_2$.
   
Now we are ready to bound both above and below $||(C_0-C_1)\pmb{\rho'}||_{tr}$, 
which produces an inequality on $\epsilon$ as will be seen soon. 
For this purpose, we first evaluate the values of $\beta$ and $k_b$ using 
geometric properties of the Bloch sphere representation of the TP-CP map on the one qubits: 
it maps the Bloch ball %(the three dimensional sphere with unit radius) 
to an ellipsoid within the Bloch ball. The proof of this technical lemma is given in Appendix.

\begin{lemma}\label{beta}
$|\beta|^2\leq \frac{1}{2}f(\epsilon)$ and $k_b\leq f(\epsilon)$ 
where $f(\epsilon)= \frac{3}{2}+\epsilon-\sqrt{\frac{9}{4}+\epsilon^2-5\epsilon}$.
\end{lemma}

Second, we decompose $||(C_0-C_1)\pmb{\rho'}||_{tr}$ as follows by the triangle inequality, 
and then bound it from above:
\begin{eqnarray}\label{bound-above}
||(C_0-C_1)\pmb{\rho'}||_{tr}
&\leq& ||(C_0-U_0)\pmb{\rho'}||_{tr} + ||U_0\pmb{\rho'}-U_0\pmb{\rho}||_{tr} 
+ ||U_1\pmb{\rho}-U_1\pmb{\rho'}||_{tr} + ||(U_1-C_1)\pmb{\rho'}||_{tr} \nonumber \\
&\leq& |\alpha|^2\cdot k_0 + ||\pmb{\rho}- \pmb{\rho'}||_{tr}\times 2 + |\alpha|^2\cdot k_1 \nonumber\\
&\leq& (k_0+k_1)|\alpha|^2 + 2|\beta|^2.  
\end{eqnarray}

Third, for the shrinking factor by the operation at $t_0$ the following lemma from \cite{NG98} is used. 
\begin{lemma}\label{NG} (Niu-Griffiths) Let $CP'_i$ be the Bloch sphere representation of $CP_i$. 
Let $l_1$ be the shortest semiaxis length of the image of $CP'_1$, 
and $l_2$ be the longest semiaxis length of the image of $CP'_2$. Then, $l_1\leq \sqrt{1-l_2^2}$.  
\end{lemma}  
Since $l_1\geq 1-2\epsilon$ by the fidelity requirement at $t_1$, Lemma \ref{NG} 
gives us the condition for $l_2$:  
\begin{equation}\label{condition-c}
l_2\leq 2\sqrt{\epsilon-\epsilon^2}. 
\end{equation}

Finally, we bound $||(C_0-C_1)\pmb{\rho'}||_{tr}$ from below by focusing on the path $s_2$-$t_2$. 
Let $M$ be the TP-CP map done at $t_2$, and $D=M(I\otimes CP_2)(I\otimes C_0-I\otimes C_1)$. 
By the fidelity requirement at $t_2$, $||D|\xi\rangle\langle\xi| ||_{tr}\geq 2-4\epsilon$ \cite{AKN98}. 
On the contrary, using the unnormalized product state $|\chi\rangle=\alpha|\psi_2\rangle|\psi_1\rangle$ 
we bound $||D|\xi\rangle\langle\xi| ||_{tr}$ by 
\[
||D|\xi\rangle\langle\xi| ||_{tr} \leq 
||D(|\xi\rangle\langle\xi|-|\chi\rangle\langle\chi|)||_{tr} + || D|\chi\rangle\langle\chi| ||_{tr}.
\]
The first term is bounded by $2 || |\xi\rangle\langle\xi|-|\chi\rangle\langle\chi| ||_{tr}$ 
since $D$ is the difference between two TP-CP maps, each of which has the operator norm at most 1 \cite{AKN98}. 
Using the monotone decreasing property of the trace distance between two states by TP-CP maps, 
the second term is bounded by 
\[
||D|\chi\rangle\langle\chi| ||_{tr} \leq 
||(I\otimes (CP_2\cdot (C_0-C_1)))|\psi_2\rangle\langle\psi_2|\otimes\pmb{\rho'} ||_{tr} 
= ||(CP_2\cdot (C_0-C_1))\pmb{\rho'} ||_{tr},
\]
which is at most $l_2 ||(C_0-C_1)\pmb{\rho'}||_{tr}$ since $CP'_2$ maps the Bloch ball 
to an ellipsoid within a ball with radius at most $l_2$. 
By a simple calculation of the trace norm, we have the following bound.
\begin{lemma}\label{just-calculation}
$|| |\xi\rangle\langle\xi| -|\chi\rangle\langle\chi| ||_{tr}\leq 2|\beta|\sqrt{1-|\beta|^2/2}.$
\end{lemma}
By Lemma \ref{just-calculation} we have 
\begin{equation}\label{bound-below}
2-4\epsilon\leq 2|| |\xi\rangle\langle\xi|-|\chi\rangle\langle\chi| ||_{tr}+ l_2 ||(C_0-C_1)\pmb{\rho'}||_{tr}
\leq 2|\beta|\sqrt{1-|\beta|^2/2} + l_2 ||(C_0-C_1)\pmb{\rho'}||_{tr}.
\end{equation}
By Lemma \ref{beta}, Ineqs.(\ref{bound-above}), (\ref{condition-c}) and (\ref{bound-below}), 
we produce an inequality on $\epsilon$ and $|\beta|$: 
\begin{equation}\label{final-ineq}
1-2\epsilon \leq 2|\beta|\sqrt{1-|\beta|^2/2}+2\sqrt{\epsilon-\epsilon^2}
\left((1-|\beta|^2)f(\epsilon)+|\beta|^2\right).
\end{equation}
(Recall that $|\alpha|^2=1-|\beta|^2$.) Note that the right-hand side of Ineq. (\ref{final-ineq}) 
is monotone increasing on $\epsilon$ and $|\beta|$ while its left-hand side is monotone decreasing on $\epsilon$. 
Therefore, by checking $\epsilon$ such that Ineq. (\ref{final-ineq}) holds under the bound of $|\beta|$ 
from Lemma \ref{beta}, we obtain $\epsilon > 1/12$.  

It still remains to prove Lemma \ref{beta}. 

\begin{proofof}{Lemma \ref{beta}} 
We only prove the bound of $|\beta|$ since the bound of $k_b$ is similarly shown. 

Let $\beta_{max}$ be the maximum of all $\beta$'s when $|\psi\rangle$ varies on the pure qubits. 
Considering the Bloch sphere representation, the point $A$ corresponding to $\pmb{\rho}$ 
is on the circle of radius $1-2|\beta_{max}|^2$ 
since $\pmb{\rho}=(1-2|\beta_{max}|^2)|\psi_1\rangle\langle\psi_1|+2|\beta_{max}|^2\cdot\frac{\pmb{I}}{2}$. 
Note that the image of the Bloch ball by the operation at $s_1$ is an ellipsoid 
whose cut $CD$ parallel to segment $AO$ has length $\leq 2-2|\beta_{max}|^2$ 
(see Fig.\ \ref{fig3}), where $O$ is the center of the Bloch ball. 
Then, by geometric properties of the ellipsoid we can show the following lemma. 

\begin{figure}
\begin{minipage}{0.5\hsize}
\begin{center}
\includegraphics*[width=5.5cm]{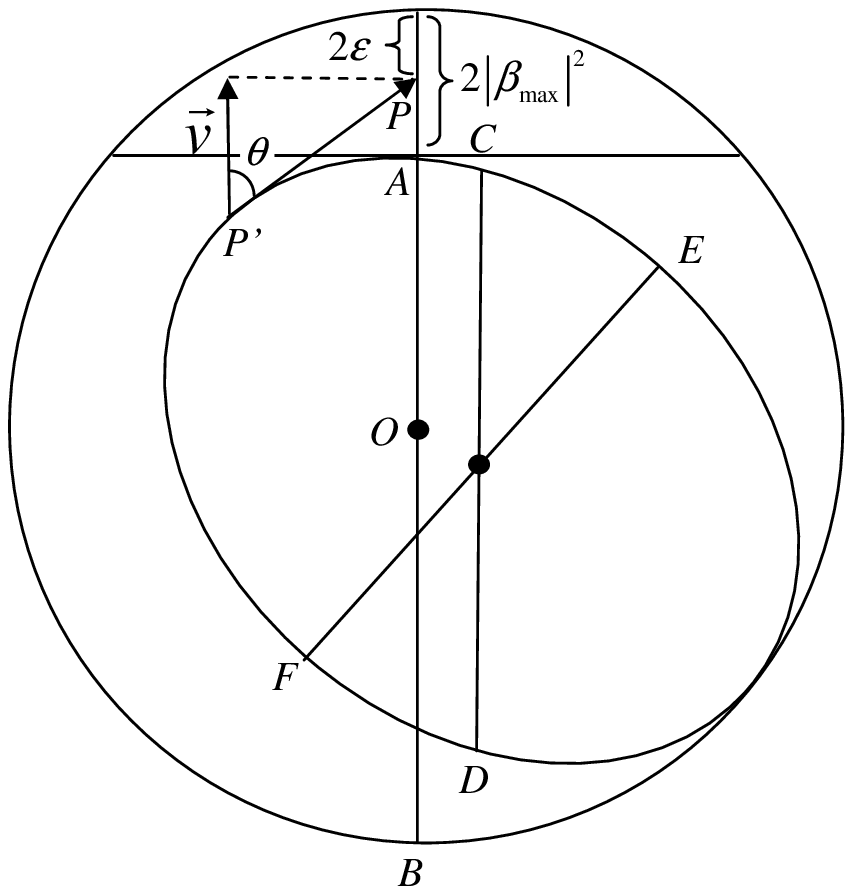}
\caption{Image by the operation at $s_1$}\label{fig3}
\end{center}
%\end{figure}
\end{minipage}\begin{minipage}{0.5\hsize}
%\begin{figure}
\begin{center}
\includegraphics*[width=7.5cm]{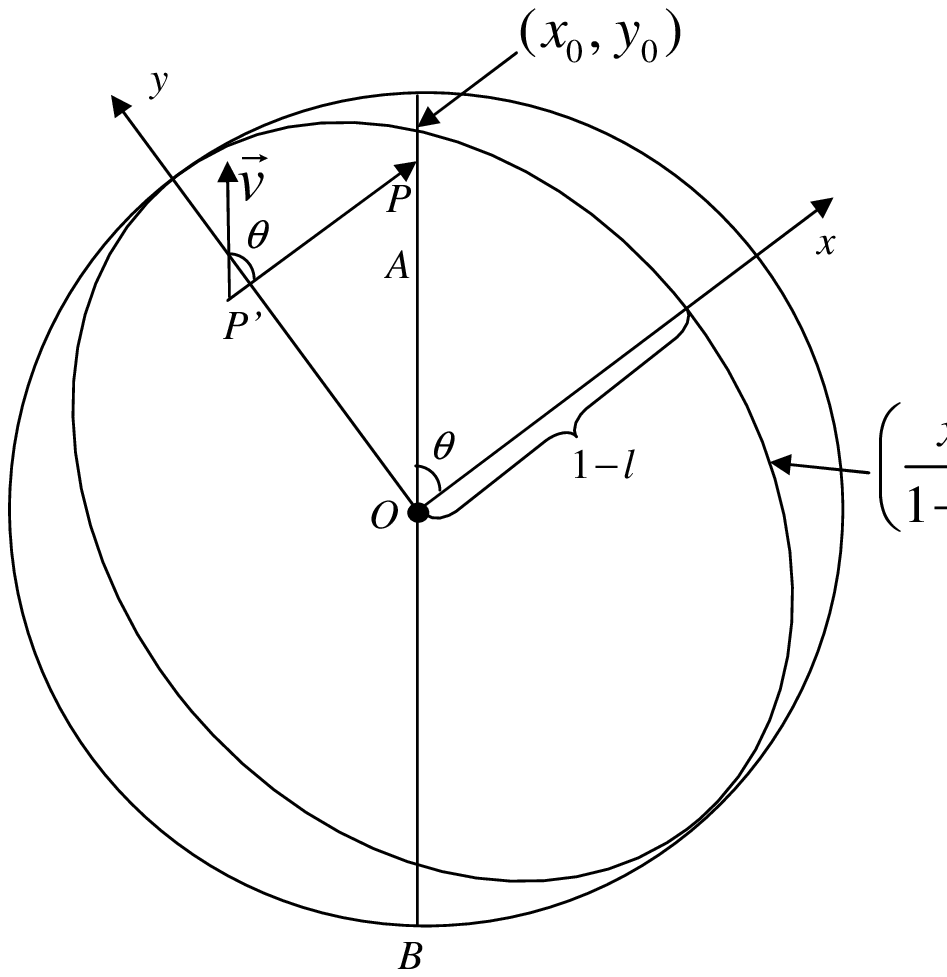}
\caption{Image by operations at $s_0$, $t_0$ and $t_1$}\label{fig4} 
\end{center}\end{minipage}
\end{figure}

\begin{claim}\label{ellipsoid}
After the operation at $t_1$, this cut is shrunk by a factor at least $1-2|\beta_{max}|^2+2\epsilon$. 
\end{claim}  

\begin{proof}
See Fig.\ \ref{fig3} again. Let $P$ be the point on the line $AB$ such that 
the length of segment $PO$ is $1-2\epsilon$. 
To satisfy the fidelity at $t_1$, some point $P'$ in the image at $s_0$ must reach $P$ 
in the final image at $t_1$. Letting $l$ be the length of $\overrightarrow{P'P}$, 
this means that the composite map by operations at $s_0$, $t_0$ and $t_1$ shrinks the Bloch ball 
by a factor of $1-l$ to the direction of $\overrightarrow{P'P}$ 
since any TP-CP map must transform the Bloch ball to an ellipsoid inside the ball.  
Let $\vec{v}$ be the component of $\overrightarrow{P'P}$ parallel to the line $AB$. 
Note that the length of $\vec{v}$ is at least $(1-2\epsilon)-(1-2|\beta_{max}|^2)=2|\beta_{max}|^2-2\epsilon$. 
Now we analyze the maximum of the shrinking factor of the Bloch ball to the direction 
of $AB$ since it is the worst case for our analysis. Since we want to know the maximum, 
we can assume that the length of $\vec{v}$ is exactly $2|\beta_{max}|^2-2\epsilon$. 
Then, the angle $\theta$ between $\overrightarrow{P'P}$ and $\vec{v}$ 
satisfies $\cos\theta=\frac{2|\beta_{max}|^2-2\epsilon}{l}$. 
Notice that the ellipse $(\frac{x}{1-l})^2+y^2=1$ (Fig.\ \ref{fig4}) is obtained by the projection 
of the final image to the plane including all the vectors we are considering. 
(Here, we assume that the plane has $x$-$y$ coordinates. Note that the $x$-axis 
is the direction parallel to $\overrightarrow{P'P}$.) 
Let $(x_0,y_0)$ be the intersection of the ellipse and the line $y=(\tan\theta)x$.  
We can see that $\sqrt{x_0^2+y_0^2}$ is the shrinking factor we want. 
By a simple calculation, this value is 
\[
\sqrt{x_0^2+y_0^2} = \sqrt{(1+\tan^2\theta)\frac{1}{(\frac{1}{1-l})^2+\tan^2\theta}}
                   = \sqrt{\frac{l(1-l)^2}{2|\beta_{max}|^2-2\epsilon +(1-l)^2 (l-2|\beta_{max}|^2+2\epsilon ) }},
\]
which is monotone decreasing as the function of $l$ (when $l\leq 1$). 
Since $l\geq 2|\beta_{max}|^2-2\epsilon$, the maximum value is obtained 
when $l=2|\beta_{max}|^2-2\epsilon$ and it is $1-2|\beta_{max}|^2+2\epsilon$. 
\end{proof}

After the operation at $t_1$, the distance between the Bloch sphere and its image by the operations 
along the $s_1$-$t_1$ line must be at most $2\epsilon$ to satisfy the fidelity requirement at $t_1$. 
Thus, the shortest semiaxis length of the final image of all transformations on the $s_1$-$t_1$ line 
must be at least $1-2\epsilon$. On the contrary, Lemma \ref{ellipsoid} implies that 
the shortest semiaxis length is at most $(1-|\beta_{max}|^2)(1-2|\beta_{max}|^2+2\epsilon)$. 
Hence, $1-2\epsilon\leq (1-|\beta_{max}|^2)(1-2|\beta_{max}|^2+2\epsilon)$, 
and we obtain $|\beta_{max}|^2\leq \frac{1}{2}\left(\frac{3}{2}+\epsilon-\sqrt{\frac{9}{4}+\epsilon^2-5\epsilon}\right)$. 
By the definition of $\beta_{max}$ we also have the same bound on $|\beta|^2$ 
for any $\beta$ corresponding to an input at $s_1$. 
\end{proofof}

Now the proof of Theorem \ref{natural-bound} is completed.
\end{proof}

\section{Protocols for Crossing Two Multiple Bits}
In this section, we consider the case that both sources are restricted 
to be one of the four $(2,1,0.85)$-quantum random access (QRA) coding states \cite{ANTV02}. 
Note that $(m,n,p)$-QRA coding is the coding of $m$ bits to $n$ qubits such that 
any one bit chosen from the $m$ bits is recovered with probability at least $p$. 
In this case, we can achieve a much better fidelity.  
As an application, we can consider a more interesting problem 
where each source node receives two classical bits, namely, $x_{1}x_{2}\in\{0,1\}^2$ at $s_1$, 
and $y_{1}y_{2}\in\{0,1\}^2$ at $s_2$. At node $t_1$, we output one 
classical bit $\mbox{Out}^1$ and similarly $\mbox{Out}^2$ at $t_2$. Now 
an adversary chooses two numbers $i_1,i_2 \in \{1,2\}$. Our protocol can 
use the information of $i_1$ only at node $t_1$ and that of $i_2$ only 
at $t_2$. Our goal is to maximize $F(x_{i_1},\mbox{Out}^1)$ and 
$F(y_{i_2},\mbox{Out}^2)$, where $F(x_{i_1},\mbox{Out}^1)$ turns out to be  
the probability that $x_{i_1} = \mbox{Out}^1$ and similarly for $F(y_{i_2},\mbox{Out}^2)$. 
Fig.~\ref{prot2} illustrates $X2C2C$ 
whose key is also on how to encode at $s_0$: it uses measurement $MM_2$ to estimate 
which QRA coding state is sent from $s_2$, and the group operation similar to $XQQ$. 
For optimal cloning at $t_0$, it uses the phase-covariant cloning \cite{Bru00,FMWI02}. 
Its details are given in Appendix. 

\begin{theorem}\label{thm-2bits}
$X2C2C$ achieves a fidelity of $1/2 + \sqrt{2}/16$ at both $t_1$ and $t_2$.
\end{theorem}

By contrast, any classical protocol cannot achieve a success probability 
greater than $1/2$ for the following reason: Let fix $y_1=y_2=0$. 
Then the path from $s_1$ to $t_1$ is obviously equivalent to 
the $(2,1,p)$-classical random access coding, where the success probability $p$ is at most $1/2$ \cite{ANTV02}. 

Extending $X2C2C$, we can also solve the above problem with probability 
$>1/2$ for the case when each source node receives three bits. The 
protocol is denoted as $X3C3C$ whose details are given in Appendix. In 
particular, $X3C3C$ needs eight different operations instead of four in 
$X2C2C$. In addition to the Pauli operations, it uses an approximation 
of the universal NOT gate \cite{BHW,GP99}, which maps a point within the Bloch sphere into its antipodes.

\begin{theorem}\label{x3c3c}
$X3C3C$ achieves a fidelity of $1/2+2/81$ at both sinks.
\end{theorem}

Interestingly, there is no $X4C4C$, which is an immediate corollary of the nonexistence 
of $(4,1,p>1/2)$-QRA coding \cite{HINRY06}. 

\begin{theorem}\label{4-1code} 
If an $X4C4C$ protocol achieves fidelity $q$, then $q\leq 1/2$. 
\end{theorem}

\begin{figure}
\begin{center}
\includegraphics*[width=5.2cm]{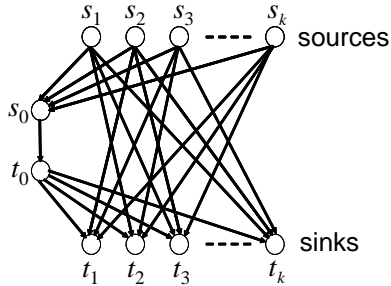}
\caption{Network $G_k$}\label{general-butterfly} 
\end{center}
%\end{minipage}
\end{figure}
 
\section{Beyond the Butterfly Network -- Concluding Remarks --}
Obviously a lot of future work remains. First of all, there is a large
gap between the current upper and lower bounds for the achievable
fidelity, which should be narrowed. Equally important is to consider
more general networks. To this direction, it might be interesting to
study the network $G_k$ as shown in Fig.~\ref{general-butterfly}, introduced in
~\cite{HKL04}. Note that there are $k$ source-sink pairs $(s_i, t_i)$
all of which share a single link from $s_0$ to $t_0$. For this network
$G_k$, we can design the protocol $XQ^k$ by a simple extension of
$XQQ$.   The idea is to decompose the node $s_0$ (similarly for $t_0$)
into a sequence of nodes of indegree two. At each of those nodes, we
do exactly the same thing as before, i.e., encoding one state by the
classical two bits obtained from the other state. It is not hard to see
that such a protocol achieves a fidelity strictly better than $1/2$. 

%%%%%%%%%%%%%%%%%%%%%
%\bibliographystyle{alpha}

%\renewcommand{\thesection}{\Roman{section}}
\appendix

\section{Appendix} \label{proof}
  
\subsection{Proofs of Lemmas in Sec.\ \ref{xqq-analysis}}

\begin{proofof}{Lemma \ref{cond}}
Consider the tetra measurement as a TP-CP map. 
For any mixed state $\pmb{\rho}$, its TP-CP map, also denoted by $TTR$, 
transforms $\pmb{\rho}$ to 
\begin{align}\label{ttr}
& \mathrm{Tr}\left[\frac{1}{2}|\chi(00)\rangle\langle\chi(00)|\pmb{\rho}\right] |00\rangle\langle 00|
+\mathrm{Tr}\left[\frac{1}{2}|\chi(01)\rangle\langle\chi(01)|\pmb{\rho}\right] |01\rangle\langle 01|\nonumber\\
&+\mathrm{Tr}\left[\frac{1}{2}|\chi(10)\rangle\langle\chi(10)|\pmb{\rho}\right] |10\rangle\langle 10|
+\mathrm{Tr}\left[\frac{1}{2}|\chi(11)\rangle\langle\chi(11)|\pmb{\rho}\right] |11\rangle\langle 11|.
\end{align}
By linearity, $TTR$ is defined by the map that transforms each 
basis matrix $|b\rangle\langle b'|$ ($b,b'=0,1$) to the matrix obtained 
by replacing $\pmb{\rho}$ in Eq.(\ref{ttr}) by $|b\rangle\langle b'|$. 
Using the definition of the states $|\chi(00)\rangle,\ldots,|\chi(11)\rangle$ we obtain 
\begin{align}\label{38-1}
TTR(|0\rangle\langle 0|)
&=\frac{1}{2}(\cos^2\tilde{\theta}(|00\rangle\langle 00|+|01\rangle\langle 01|)
            +\sin^2\tilde{\theta}(|10\rangle\langle 10|+|11\rangle\langle 11|))\nonumber\\
&=2\sin^2\tilde{\theta}\left(\frac{\pmb{I}}{2}\otimes\frac{\pmb{I}}{2}\right)
 +(\cos^2\tilde{\theta}-\sin^2\tilde{\theta})\left(|0\rangle\langle 0|\otimes\frac{\pmb{I}}{2}\right)\nonumber\\
&=\frac{1}{\sqrt{3}}\left(|0\rangle\langle 0|\otimes\frac{\pmb{I}}{2}\right)
 +\left(1-\frac{1}{\sqrt{3}}\right)\left(\frac{\pmb{I}}{2}\otimes\frac{\pmb{I}}{2}\right).
\end{align}
Similarly, we can check that 
\begin{align}
TTR(|1\rangle\langle 1|)&=\frac{1}{\sqrt{3}}\left(|1\rangle\langle 1|\otimes\frac{\pmb{I}}{2}\right)
 +\left(1-\frac{1}{\sqrt{3}}\right)\left(\frac{\pmb{I}}{2}\otimes\frac{\pmb{I}}{2}\right),\label{38-2}\\
TTR(|0\rangle\langle 1|)&=
\frac{1}{4\sqrt{3}}\left(
(|00\rangle\langle 00|+|10\rangle\langle 10|-|11\rangle\langle 11|-|01\rangle\langle 01|)\right.\nonumber\\
&\left. +\imath
(|00\rangle\langle 00|+|11\rangle\langle 11|-|01\rangle\langle 01|-|10\rangle\langle 10|)
\right)\label{38-3} \\
TTR(|1\rangle\langle 0|)&=TTR(|0\rangle\langle 1|)^\dagger.\label{38-4}
\end{align}
Now we consider the TP-CP map of $GR({\cal Q},TTR({\cal Q}'))$. 
Assume that the state of ${\cal Q}$ is $\pmb{\rho}$. 
Recall that $GR(\pmb{\rho},|00\rangle\langle 00|)=\pmb{\rho}$, 
$GR(\pmb{\rho},|01\rangle\langle 01|)=Z\pmb{\rho}$, $GR(\pmb{\rho},|10\rangle\langle 10|)=X\pmb{\rho}$, 
and $GR(\pmb{\rho},|11\rangle\langle 11|)=Y\pmb{\rho}$.  
If the GR operation under the two bits being selected uniformly at random, 
which means that the state of $TTR({\cal Q}')$ is the state $\frac{\pmb{I}}{2}\otimes\frac{\pmb{I}}{2}$, 
is applied to $\pmb{\rho}$, then $\pmb{\rho}$ is mapped to $\frac{\pmb{I}}{2}$ 
since the four GR operations are evenly applied to $\pmb{\rho}$. 
Thus, Eqs.(\ref{38-1}) and (\ref{38-2}) imply that 
$GR(\cdot,TTR(\cdot))$ maps $\pmb{\rho}\otimes |0\rangle\langle 0|$ 
and $\pmb{\rho}\otimes |1\rangle\langle 1|$ 
to $\frac{1}{\sqrt{3}}\frac{I\pmb{\rho}+Z\pmb{\rho}}{2} +(1-\frac{1}{\sqrt{3}})\frac{\pmb{I}}{2}$ 
and $\frac{1}{\sqrt{3}}\frac{X\pmb{\rho}+Y\pmb{\rho}}{2} +(1-\frac{1}{\sqrt{3}})\frac{\pmb{I}}{2}$, 
respectively. Moreover, by Eq.(\ref{38-3}) $\pmb{\rho}\otimes |0\rangle\langle 1|$ is mapped 
to 
$$
\frac{1}{4\sqrt{3}}( (I\pmb{\rho}+X\pmb{\rho})-(Y\pmb{\rho}+Z\pmb{\rho}) 
+ \imath ( (I\pmb{\rho}+Y\pmb{\rho})-(Z\pmb{\rho}+X\pmb{\rho}) ) ),
$$
which is $\frac{1}{2\sqrt{3}}(V(I,X)\pmb{\rho}-V(Y,Z)\pmb{\rho}+\imath(V(I,Y)\pmb{\rho}-V(Z,X)\pmb{\rho}))$. 
%%Since $TTR(|1\rangle\langle 0|)=TTR(|0\rangle\langle 1|)^\dagger$, 
By Eq.(\ref{38-4}) it is clear that $\pmb{\rho}\otimes |1\rangle\langle 0|$ is mapped to 
$\frac{1}{2\sqrt{3}}(V(I,X)\pmb{\rho}-V(Y,Z)\pmb{\rho}-\imath(V(I,Y)\pmb{\rho}-V(Z,X)\pmb{\rho}))$. 

Next, we show that $V(I,Z)$, which maps $\pmb{\rho}$ to $\frac{1}{2}(I\pmb{\rho}+Z\pmb{\rho})$, 
is an $\pmb{I}$-invariant TP-CP map (similarly shown for the other five operations). 
In fact, $V(I,Z)$ is a TP-CP map since it is implementable by the following operation: 
Choose a bit $r$ uniformly at random, and apply $Z$ to $\pmb{\rho}$ if $r=1$. 
Its $\pmb{I}$-invariance comes from the fact that the Pauli operations are $\pmb{I}$-invariant.   
\end{proofof}

\begin{proofof}{Lemma \ref{simplify}}
Let $\pmb{\rho}=\left(
\begin{array}{ll}
a & b\\
c & d
\end{array}
\right)$. Then, we can check that 
$$
Z\pmb{\rho}=\left(
\begin{array}{ll}
a & -b\\
-c & d
\end{array}
\right),\ \   
X\pmb{\rho}= \left(
\begin{array}{ll}
d & c\\
b & a
\end{array}
\right),\ \  
Y\pmb{\rho}= \left(
\begin{array}{ll}
d & -c\\
-b & a
\end{array}
\right).
$$
Thus, $V(I,Z)$ is the TP-CP map that maps $\pmb{\rho}$ 
to $V(I,Z)\pmb{\rho}=\left(
\begin{array}{ll}
a & 0\\
0 & d
\end{array}
\right)$. Similarly, we can see that $V(X,Y)$, $V(I,X)$, $V(Y,Z)$, $V(I,Y)$, 
and $V(Z,X)$ are TP-CP maps that map $\pmb{\rho}$ to $V(X,Y)\pmb{\rho}=\left(
\begin{array}{ll}
d & 0\\
0 & a
\end{array}
\right)$, $V(I,X)\pmb{\rho}=\left(\begin{array}{ll}
\frac{1}{2} & \frac{b+c}{2} \\
\frac{b+c}{2} & \frac{1}{2}
\end{array}
\right)$, $V(Y,Z)\pmb{\rho}=\left(
\begin{array}{ll}
\frac{1}{2} & -\frac{b+c}{2}\\
-\frac{b+c}{2} & \frac{1}{2}
\end{array}
\right)$, 
$V(I,Y)\pmb{\rho}=\left(
\begin{array}{ll}
\frac{1}{2} & \frac{b-c}{2}\\
\frac{c-b}{2} & \frac{1}{2}
\end{array}
\right)$, and 
$V(Z,X)\pmb{\rho}=\left(
\begin{array}{ll}
\frac{1}{2} & \frac{c-b}{2}\\
\frac{b-c}{2} & \frac{1}{2}
\end{array}
\right)$, respectively. Using these TP-CP maps, we can check that 1)-7) hold.  
\end{proofof}

\

\noindent
{\bf Calculation of $\pmb{\rho}_{out}^1$.}  
Here, we give the calculation to obtain $\pmb{\rho}_{out}^1$ in Sec.\ \ref{xqq-analysis}. 
Recall the state on ${\cal Q}_5\otimes{\cal Q}_4$ (Eq.\ (\ref{formula})). 
Using Lemma \ref{simplify}(7) and Lemma \ref{cond}, the state $\pmb{\rho}_{out}^1$ is represented as 
\begin{align}\label{formula2}
&\left(\frac{1}{\sqrt{3}}\right)^2
\left(\frac{2e}{3}V(I,Z)V(I,Z)\pmb{\rho}_1+\frac{1}{6}V(X,Y)V(I,Z)\pmb{\rho}_1\right)\nonumber\\
&+\left(\frac{1}{2\sqrt{3}}\right)^2\frac{1}{6}
V(I,X;I,Y;-)V(I,X;I,Y;+)\pmb{\rho}_1+\left(\frac{1}{2\sqrt{3}}\right)^2\frac{1}{6}
V(I,X;I,Y;+)V(I,X;I,Y;-)\pmb{\rho}_1\nonumber\\
&+\left(\frac{1}{\sqrt{3}}\right)^2
\left(\frac{2h}{3}V(X,Y)V(X,Y)\pmb{\rho}_1+\frac{1}{6}V(I,Z)V(X,Y)\pmb{\rho}_1\right),
\end{align} 
where the terms $\frac{\pmb{I}}{2}$ produced from the second, 
fourth, sixth, and eighth terms of Eq.(\ref{formula}) by Lemma \ref{cond}(7) are omitted. 
Moreover, Eq.(\ref{formula2}) is rewritten as follows using Lemma \ref{cond}(1-2) 
for the first and last terms of Eq.(\ref{formula2}) and Lemma \ref{cond}(3-7) 
for the second and third terms of Eq.(\ref{formula2}): 
\begin{align*}
&\frac{1}{3}\left[\frac{2e+2h}{3}\left(
\begin{array}{ll}
a & 0\\
0 & d
\end{array}
\right)
+
\frac{1}{3}
\left(
\begin{array}{ll}
d & 0\\
0 & a
\end{array}
\right)
\right]
+
\frac{1}{12}\left[\frac{1}{6}\left(
\begin{array}{ll}
\frac{1}{2} & \frac{b+c}{2}\\
\frac{b+c}{2} & \frac{1}{2}
\end{array}
\right)\times 4
-
\frac{1}{6}\left(
\begin{array}{ll}
\frac{1}{2} & -\frac{b+c}{2}\\
-\frac{b+c}{2} & \frac{1}{2}
\end{array}
\right)\times 4
\right]\\
&+
\frac{1}{12}\left[\frac{1}{6}\left(
\begin{array}{ll}
\frac{1}{2} & \frac{b-c}{2}\\
\frac{c-b}{2} & \frac{1}{2}
\end{array}
\right)\times 4
-
\frac{1}{6}\left(
\begin{array}{ll}
\frac{1}{2} & \frac{c-b}{2}\\
\frac{b-c}{2} & \frac{1}{2}
\end{array}
\right)\times 4
\right],
\end{align*}
where the terms $\frac{\pmb{I}}{2}$ produced by Lemma \ref{cond} are omitted. 
Using $e+h=a+d=1$, the above formula is finally rewritten as $\frac{1}{9}\pmb{\rho}_1+\frac{1}{9}\pmb{I}$. 

\subsection{Computing $F(|\psi_2\rangle,\pmb{\rho}_{out}^2)$}\label{s2t2}
We investigate the quality of the path from $s_2$ to $t_2$. 
Fix $\pmb{\rho}_1=|\psi_1\rangle\langle\psi_1|$ as an arbitrary state at node $s_1$, 
and consider four maps $D_1$: $|\psi_2\rangle\rightarrow {\cal Q}_3$, $D_2[\pmb{\rho}_1]$: 
${\cal Q}_3\rightarrow{\cal Q}_5$, $D_3$: ${\cal Q}_5\rightarrow{\cal Q}_6$ 
and $D_4[\pmb{\rho}_1]$: ${\cal Q}_6\rightarrow \pmb{\rho}_{out}^2$. 
We wish to compute the fidelity of the composite map 
$D_{s_2t_2}=D_4[\pmb{\rho}_1]\circ D_3\circ D_2[\pmb{\rho}_1]\circ D_1$. 

\begin{lemma}\label{c3c2-dash}
$D_3\circ D_2[\pmb{\rho}_1]= D_2[\pmb{\rho}_1]\circ D_3$. 
\end{lemma}

\begin{proof}
As with Lemma \ref{c3c2}, it suffices to show that $C^{(0)}$ is commutative with $D_2[\pmb{\rho}_1]$. 
For this purpose, we prove that, for any qubit $\pmb{\rho}$ and 
any matrix basis element $|b\rangle\langle b'|$ (where $b,b'\in\{0,1\}$) on ${\cal Q}_2$, 
$C^{(0)}(GR(|b\rangle\langle b'|,TTR(\pmb{\rho})))=GR(|b\rangle\langle b'|,TTR(C^{(0)}(\pmb{\rho})))$. 
We first evaluate the left-hand side. Let $T_{00}$ (resp.\ $T_{01}$, $T_{10}$, $T_{11}$) 
be the map that transforms $\pmb{\rho}$ to $I\pmb{\rho}I^\dagger$ 
(resp.\ $Z\pmb{\rho}Z^\dagger$, $X\pmb{\rho}X^\dagger$, $Y\pmb{\rho}Y^\dagger$).  
Since $C^{(0)}$ is $0$-shrinking, 
\begin{align*}
C^{(0)}(GR(|b\rangle\langle b'|,TTR(\pmb{\rho}))) &=C^{(0)}\left(
\sum_{r\in\{0,1\}^2}\langle r|TTR(\pmb{\rho})|r \rangle T_r(|b\rangle\langle b'|)
\right)\\
&=\sum_{ r\in\{0,1\}^2 }\langle r|TTR(\pmb{\rho})|r \rangle T_r\circ C^{(0)}
(|b\rangle\langle b'|)=
\left\{
\begin{array}{ll}
\pmb{O} & (\mathrm{ if }\ b\neq b')\\
\frac{\pmb{I}}{2} & (\mathrm{ if }\ b=b').
\end{array}
\right.
\end{align*}
Here, we used the commutativity between $C^{(0)}$ and the group operations for the second equality. 
Next, we evaluate the right-hand side:
\begin{align*}
GR(|b\rangle\langle b'|,TTR(C^{(0)}(\pmb{\rho})))
&=\sum_{r\in\{0,1\}^2}\langle r|\frac{\pmb{I}}{2}\otimes\frac{\pmb{I}}{2}|r\rangle T_r(|b\rangle\langle b'|)\\
&=\frac{1}{4}\sum_{r\in\{0,1\}^2} T_r(|b\rangle\langle b'|)
=
\left\{
\begin{array}{ll}
\pmb{O} & (\mathrm{ if }\ b\neq b')\\
\frac{\pmb{I}}{2} & (\mathrm{ if }\ b=b').
\end{array}
\right.
\end{align*}
Here, the first equality is obtained since the tetra measurement for $\frac{\pmb{I}}{2}$ 
gives two bits uniformly at random, and the third equality is obtained 
since $I|b\rangle\langle b'|I^\dagger+Z|b\rangle\langle b'|Z^\dagger+X|b\rangle\langle b'|X^\dagger
+Y|b\rangle\langle b'|Y^\dagger$ is $2\pmb{I}$ if $b=b'$, and $\pmb{O}$ otherwise.  
\end{proof}

\begin{lemma}\label{mainlemma2}
(Main Lemma) $D_4[\pmb{\rho}_1]\circ D_2[\pmb{\rho}_1]$ is $\pmb{I}$-invariant 
and for any pure state $|\psi_2\rangle$, 
$F(|\psi_2\rangle,D_4[\pmb{\rho}_1]\circ D_2[\pmb{\rho}_1](|\psi_2\rangle))=1/2+\sqrt{3}/54$.
\end{lemma}

\begin{lemma}\label{final-2}
For any pure state $|\psi_2\rangle$, $F(|\psi_2\rangle,D_{s_2t_2}(|\psi_2\rangle))=1/2+2\sqrt{3}/243$.
\end{lemma}

\begin{proof}
By Lemma \ref{c3c2-dash}, $D_{s_2t_2}=D_4[\pmb{\rho}_1]\circ D_2[\pmb{\rho}_1]\circ D_3\circ D_1$. 
By Lemmas \ref{shrink-composition} and \ref{shrink}, 
$D_3\circ D_1(|\psi_1\rangle)=\frac{4}{9}|\psi_2\rangle\langle\psi_2|+\frac{5}{9}\cdot\frac{\pmb{I}}{2}$. 
By Lemma \ref{mainlemma2}, we have $D_{s_2t_2}(|\psi_2\rangle)
=D_4[\pmb{\rho}_1]\circ D_2[\pmb{\rho}_1](\frac{4}{9}|\psi_2\rangle\langle\psi_2|
+\frac{5}{9}\cdot\frac{\pmb{I}}{2})=\frac{4}{9}\pmb{\rho}+\frac{5}{9}\cdot\frac{\pmb{I}}{2}$ 
for some $\pmb{\rho}$ such that $F(|\psi_2\rangle,\pmb{\rho})=\frac{1}{2}+\frac{\sqrt{3}}{54}$. Thus, 
$F(|\psi_2\rangle,D_{s_2t_2}(|\psi_2\rangle))=\frac{4}{9}(\frac{1}{2}+\frac{\sqrt{3}}{54})
+\frac{5}{9}\cdot\frac{1}{2}=\frac{1}{2}+\frac{2\sqrt{3}}{243}$.
\end{proof}

\begin{proofof}{Lemma \ref{mainlemma2}} 
It is not hard to see that $D_4[\pmb{\rho}_1]\circ D_2[\pmb{\rho}_1]$ is $\pmb{I}$-invariant 
from the following description. To compute $D_4[\pmb{\rho}_1]\circ D_2[\pmb{\rho}_1](|\psi_2\rangle)$, 
let $|\psi_2\rangle=\cos\theta_1|0\rangle+e^{\imath\theta_2}\sin\theta_1|1\rangle$ be the state 
on ${\cal Q}_3$, $\pmb{\rho}_1=\left(\begin{array}{ll}
a & b\\c & d
\end{array}
\right)$, and ${\cal Q}_5={\cal Q}_6$. Let $p[r_1r_2]$ be the probability 
that $r_1r_2$ is obtained by the tetra measurement of ${\cal Q}_3$. Then, we have the following values.  
\begin{lemma}\label{tetra-probability}
\begin{align*}&p[00]=\frac{1}{4}+\frac{\sqrt{3}}{12}
\left(\cos 2\theta_1+\sin 2\theta_1(\cos\theta_2+\sin\theta_2)\right),\\ 
&p[01]=\frac{1}{4}+\frac{\sqrt{3}}{12}
\left(\cos 2\theta_1+\sin 2\theta_1(-\cos\theta_2-\sin\theta_2)\right),\\
&p[10]=\frac{1}{4}+\frac{\sqrt{3}}{12}
\left(-\cos 2\theta_1+\sin 2\theta_1(\cos\theta_2-\sin\theta_2)\right),\\ 
&p[11]=\frac{1}{4}+\frac{\sqrt{3}}{12}
\left(-\cos 2\theta_1+\sin 2\theta_1(-\cos\theta_2+\sin\theta_2)\right).
\end{align*}
\end{lemma}

\begin{proof}%of}{Lemma \ref{tetra-probability}}
Let $\rho_2=\left(\begin{array}{ll}
e & f\\ f^* & h\end{array}\right)$. Then,  
\begin{align*}
p[00]&=\mathrm{Tr}\left[\frac{1}{2}|\chi(00)\rangle\langle\chi(00)| \left(
\begin{array}{ll}
e & f\\f^* & h
\end{array}
\right)
\right]\\
&=\frac{1}{2}(e\cos^2\tilde{\theta}+h\sin^2\tilde{\theta}+f\sin\tilde{\theta}\cos\tilde{\theta}e^{\imath\pi/4}
+f^*\sin\tilde{\theta}\cos\tilde{\theta}e^{-\imath\pi/4}).
\end{align*}
Similarly, we obtain
\begin{align*}
&p[01]=\frac{1}{2}
(e\cos^2\tilde{\theta}+h\sin^2\tilde{\theta}
+f\sin\tilde{\theta}\cos\tilde{\theta}e^{-3\imath\pi/4}
+f^*\sin\tilde{\theta}\cos\tilde{\theta}e^{3\imath\pi/4}),\\
&p[10]=\frac{1}{2}
(e\sin^2\tilde{\theta}+h\cos^2\tilde{\theta}
+f\sin\tilde{\theta}\cos\tilde{\theta}e^{-\imath\pi/4}
+f^*\sin\tilde{\theta}\cos\tilde{\theta}e^{\imath\pi/4}),\\
&p[11]=\frac{1}{2}
(e\sin^2\tilde{\theta}+h\cos^2\tilde{\theta}
+f\sin\tilde{\theta}\cos\tilde{\theta}e^{3\imath\pi/4}
+f^*\sin\tilde{\theta}\cos\tilde{\theta}e^{-3\imath\pi/4}).
\end{align*}
Because $e=\cos^2\theta_1$, $h=\sin^2\theta_1$, $f=e^{-\imath\theta_2}\sin\theta_1\cos\theta_1$, 
and $\cos^2\tilde{\theta}=1/2+\sqrt{3}/6$ (and hence $\sin\tilde{\theta}\cos\tilde{\theta}=1/\sqrt{6}$), 
$p[00]$ is rewritten as
\begin{align*}
p[00]
&=\frac{1}{2}\left(
\frac{1}{2}+\frac{\sqrt{3}}{6}(\cos^2\theta_1-\sin^2\theta_1)
+\frac{1}{\sqrt{6}}\sin\theta_1\cos\theta_1
\left(e^{-\imath\theta_2}\frac{1+\imath}{\sqrt{2}}
+e^{\imath\theta_2}\frac{1-\imath}{\sqrt{2}}
\right)
\right)
\\
&=\frac{1}{4}+\frac{\sqrt{3}}{12}
(\cos 2\theta_1 +\sin 2\theta_1 (\cos\theta_2+\sin\theta_2 ) ),
\end{align*}
which is the desired value. Similarly, we can calculate the desired values of $p[01],p[10],p[11]$.  
\end{proof}%of}

Then, after the group operation, the state on ${\cal Q}_1\otimes{\cal Q}_5$ 
$={\cal Q}_1\otimes{\cal Q}_6$ can be written as 
$$\pmb{\sigma}=p[00]\pmb{\rho}(I)+p[01]\pmb{\rho}(Z)+p[10]\pmb{\rho}(X)+p[11]\pmb{\rho}(Y),
$$
where $\pmb{\rho}(I),\ldots$ can be given by the following lemma:
\begin{lemma}\label{q1q6}
For $W\in\{I,X,Y,Z\}$, 
\begin{eqnarray*}
\pmb{\rho}(W)
&=&\left(
\frac{2a}{3}|0\rangle\langle 0|+\frac{1}{6}|1\rangle\langle 1|
+\frac{b}{3}|0\rangle\langle 1|+\frac{c}{3}|1\rangle\langle 0|
\right)\otimes W\left(|0\rangle\langle 0|\right)\\
&+& 
\left(
\frac{1}{6}|1\rangle\langle 0|+\frac{b}{3}I\right)\otimes W(|0\rangle\langle 1|)
   +\left(
\frac{1}{6}|0\rangle\langle 1|+\frac{c}{3}I
\right)\otimes W(|1\rangle\langle 0|) \\
&+&\left(
\frac{1}{6}|0\rangle\langle 0|+\frac{2d}{3}|1\rangle\langle 1|
+\frac{b}{3}|0\rangle\langle 1|+\frac{c}{3}|1\rangle\langle 0|
\right)\otimes  
W\left(|1\rangle\langle 1|\right).  
\end{eqnarray*}
\end{lemma}

\begin{proof}%of}{Lemma \ref{q1q6}}
By Eq.(\ref{BHcopy}), we can see that the state of the system ${\cal Q}_1\otimes{\cal Q}_2$ is
\begin{align*}
&\left(\frac{2a}{3}|0\rangle\langle 0|+\frac{b}{3}|0\rangle\langle 1|
+\frac{c}{3}|1\rangle\langle 0|+\frac{1}{6}|1\rangle\langle 1|
\right)\otimes |0\rangle\langle 0|
+\left(\frac{1}{6}|1\rangle\langle 0|+\frac{b}{3}\pmb{I}\right)\otimes |0\rangle\langle 1|\\
&+ \left(\frac{1}{6}|0\rangle\langle 1|+\frac{c}{3}\pmb{I}\right)\otimes |1\rangle\langle 0|
+\left(\frac{1}{6}|0\rangle\langle 0|+\frac{b}{3}|0\rangle\langle 1|
+\frac{c}{3}|1\rangle\langle 0|+\frac{2d}{3}|1\rangle\langle 1|
\right)\otimes |1\rangle\langle 1|,
\end{align*}
and hence we obtain the desired formula. 
\end{proof}%of}

Then, we apply the 3D Bell measurement to $\pmb{\sigma}$, obtaining $\pmb{\rho}_{out}^2=BM(\pmb{\sigma})$. 
For this calculation, we need the following lemma: 
 
\begin{lemma}\label{bell}
1) If $W\in\{I,Z\}$ ($\in\{X,Y\}$, resp.), 
then by operation (a) we obtain the state $\frac{1}{3}|0\rangle\langle 0|+\frac{2}{3}\cdot\frac{\pmb{I}}{2}$ 
($\frac{1}{3}|1\rangle\langle 1|+\frac{2}{3}\cdot\frac{\pmb{I}}{2}$, resp.) as $BM(\pmb{\rho}(W))$. 

2) If $W\in\{I,X\}$ ($\in\{Y,Z\}$, resp.), 
then by operation (b) we obtain the state $\frac{1}{3}|+\rangle\langle +|+\frac{2}{3}\cdot\frac{\pmb{I}}{2}$ 
($\frac{1}{3}|-\rangle\langle -|+\frac{2}{3}\cdot\frac{\pmb{I}}{2}$, resp.) as $BM(\pmb{\rho}(W))$. 

3) If $W\in\{I,Y\}$ ($\in\{Z,X\}$, resp.), 
then by operation (c) we obtain the state $\frac{1}{3}|+'\rangle\langle +'|+\frac{2}{3}\cdot\frac{\pmb{I}}{2}$ 
($\frac{1}{3}|-'\rangle\langle -'|+\frac{2}{3}\cdot\frac{\pmb{I}}{2}$, resp.) as $BM(\pmb{\rho}(W_1,W_2))$. 
\end{lemma}

\begin{proof}%of}{Lemma \ref{bell}}
By calculating the probabilities that the measurement values for the four Bell states are obtained, 
we have Table \ref{table2}. (This is checked by calculating 
$\langle\Phi^+|\pmb{\rho}(W)|\Phi^+\rangle$, 
$\langle\Phi^-|\pmb{\rho}(W)|\Phi^-\rangle$,
$\langle\Psi^+|\pmb{\rho}(W)|\Psi^+\rangle$, and 
$\langle\Psi^-|\pmb{\rho}(W)|\Psi^-\rangle$ for each $W\in\{I,X,Y,Z\}$.) 
According to Table \ref{table2}, we can obtain the desired result 
by a simple calculation. For example, in case where $W=Z$ is applied 
at $s_0$, by operation (a) we obtain the state 
$(\frac{1}{3}+\frac{1}{3})|0\rangle\langle 0|+(0 +\frac{1}{3})|1\rangle\langle 1|
=\frac{1}{3}|0\rangle\langle 0|+\frac{2}{3}\cdot\frac{\pmb{I}}{2}$. 
The other cases are similarly checked. 
\begin{table}[tb]
\begin{center}
\begin{tabular}{|l|l|l|l|l|} \hline
Operation$\backslash$Bell states & $|\Phi^+\rangle$ & $|\Phi^-\rangle$ & $|\Psi^+\rangle$ & $|\Psi^-\rangle$ \\ \hline
$I$       & $1/3$  & $1/3$  & $1/3$  & $0$   \\ \hline
$Z$       & $1/3$  & $1/3$  & $0$   & $1/3$   \\ \hline
$X$       & $1/3$  & $0$   & $1/3$  & $1/3$   \\ \hline
$Y$       & $0$   & $1/3$  & $1/3$  & $1/3$   \\ \hline 
\end{tabular}
\caption{Probabilities that the measurement values for the four Bell states 
by the measurement in the Bell basis at $t_2$ are obtained according to  
the approximated group operations at $s_0$.}  \label{table2}
\end{center}
\end{table}
\end{proof}%of}
 
Now we can compute $\pmb{\rho}_{out}^2=BM(\pmb{\sigma})=p[00] BM(\pmb{\rho}(I))+\cdots$ 
by summing up (1)-(3) of Lemma \ref{bell} with weight $1/3$ for each, 
which implies $\pmb{\rho}_{out}^2=\frac{2}{3}\cdot\frac{\pmb{I}}{2}+\frac{1}{9}\pmb{\rho}_0$, where 
\begin{eqnarray*}
\pmb{\rho}_0
&=& 
p[00](|0\rangle\langle 0|+|+\rangle\langle +|+|+'\rangle\langle +'|)
+ 
p[01](|0\rangle\langle 0|+|-\rangle\langle -|+|-'\rangle\langle -'|)\\
&+&
p[10](|1\rangle\langle 1|+|+\rangle\langle +|+|-'\rangle\langle -'|)
+ 
p[11](|1\rangle\langle 1|+|-\rangle\langle -|+|+'\rangle\langle +'|).
\end{eqnarray*}

We can check the following lemma by a simple calculation. 

\begin{lemma}\label{calculation}
$\langle\psi_2|\pmb{\rho}_0|\psi_2\rangle = 3/2+\sqrt{3}/6$.
\end{lemma}

\begin{proof}%of}{Lemma \ref{calculation}}
By the definition of $p[r_1r_2]$, $\pmb{\rho}_0=3/2+\sqrt{3}\pmb{\rho}'_0/12$, where
\begin{eqnarray*}
\pmb{\rho}'_0
&=& 
(\cos 2\theta_1+\sin 2\theta_1(\cos\theta_2+\sin\theta_2) )
(|0\rangle\langle 0|+|+\rangle\langle +|+|+'\rangle\langle +'|)\\
&+& 
(\cos 2\theta_1+\sin 2\theta_1(-\cos\theta_2-\sin\theta_2) )
(|0\rangle\langle 0|+|-\rangle\langle -|+|-'\rangle\langle -'|)\\
&+&
(-\cos 2\theta_1+\sin 2\theta_1(\cos\theta_2-\sin\theta_2) )
(|1\rangle\langle 1|+|+\rangle\langle +|+|-'\rangle\langle -'|)\\
&+& 
(-\cos 2\theta_1+\sin 2\theta_1(-\cos\theta_2+\sin\theta_2) )
(|1\rangle\langle 1|+|-\rangle\langle -|+|+'\rangle\langle +'|).
\end{eqnarray*}
Thus, it suffices to show that $\langle\psi_2|\pmb{\rho}'_0|\psi_2\rangle =2$. 
We can rewrite $\pmb{\rho}'_0$ as 
\begin{align*}
\pmb{\rho}'_0
&=2\cos 2\theta_1(|0\rangle\langle 0|-|1\rangle\langle 1|)
+2\sin 2\theta_1\cos\theta_2(|+\rangle\langle +|-|-\rangle\langle -|)\\
&+2\sin 2\theta_1\sin\theta_2(|+'\rangle\langle +'|-|-'\rangle\langle -'|).
\end{align*}
Recalling that $|\psi_2\rangle=\cos\theta_1|0\rangle+e^{\imath\theta_2}\sin\theta|1\rangle$, 
we can check that $\langle\psi_2||0\rangle\langle 0|-|1\rangle\langle 1||\psi_2\rangle=\cos 2\theta_1$, 
$\langle\psi_2||+ \rangle\langle + | -|-\rangle\langle -  ||\psi_2\rangle =\sin 2\theta_1\cos\theta_2$, 
and $\langle\psi_2||+'\rangle\langle +'|- |-'\rangle\langle -'||\psi_2\rangle =\sin 2\theta_1\sin\theta_2$.  
Thus, we obtain $
\langle\psi_2|\pmb{\rho}'_0|\psi_2\rangle 
=2\cos^2 2\theta_1+2\sin^2 2\theta_1\cos^2\theta_2+2\sin^2 2\theta_1\sin^2\theta_2=2$.
\end{proof}%of}

By Lemma \ref{calculation}, we finally obtain $\langle\psi_2|\pmb{\rho}_{out}^2|\psi_2\rangle 
=\frac{1}{3}+\frac{1}{9}\left(\frac{3}{2}+\frac{\sqrt{3}}{6}\right)
= \frac{1}{2}+\frac{\sqrt{3}}{54}$. This completes the proof of Lemma \ref{mainlemma2}. 
\end{proofof}

\subsection{Proof of Theorem \ref{xqq}}
Our bound is given under the case where the sources at $s_1$ and $s_2$ 
are a qubit $|\psi\rangle$ and a classical bit $b$. Also, we assume 
that two side links have unlimited capacity, and then we can assume that 
the encoded states from sources are pure states. 
Suppose that there is a protocol with fidelity $1-\epsilon$. Then, we show $\epsilon > 0.017$. 
Let $|\xi_\psi\rangle_{t_2s_0}$ be the encoded state sent from $s_1$, 
and $|\phi(b)\rangle_{s_0t_1}$ be the encoded state sent from $s_2$, 
where the subscript of the ket vector presents where they are in. 
By the Schmidt decomposition, 
they are written as: $|\xi_\psi\rangle_{t_2s_0}=\alpha|\psi_2\rangle_{t_2}|\psi_1\rangle_{s_0}
+\beta|\psi_2^\bot\rangle_{t_2}|\psi_1^\bot\rangle_{s_0}$, 
and $|\phi(b)\rangle_{s_0t_1}=\gamma_b|\phi(b)_1\rangle_{s_0}|\phi(b)_2\rangle_{t_1} 
+\delta_b|\phi(b)_1^\bot\rangle_{s_0}|\phi(b)_2^\bot\rangle_{t_1}$. 
Without loss of generality, $|\beta|\leq|\alpha|$ and $|\delta_b|\leq|\gamma_b|$. 
Note that $\pmb{\rho}_\psi = |\alpha|^2|\psi_1\rangle\langle\psi_1|
+|\beta|^2|\psi_1^\bot\rangle\langle\psi_1^\bot|$ (and 
$\pmb{\rho}(b)=|\gamma|^2|\phi(b)_1\rangle\langle\phi(b)_1|
+|\delta|^2|\phi(b)_1^\bot\rangle\langle\phi(b)_1^\bot|$, resp.) are the states 
after the operations at $s_1$ (and $s_2$, resp.) when we focus on the path from $s_1$ to $t_1$ 
(the path from $s_2$ to $t_2$, resp.). Then, we have the following bounds on $\beta$ and $\delta_b$.

\begin{lemma}\label{beta-delta} 
$|\beta|^2$ and $|\delta_b|^2$ are at most 
$\frac{1}{2}\left(\frac{3}{2}+\epsilon-\sqrt{\frac{9}{4}+\epsilon^2-5\epsilon}\right)$, 
and $|\delta_0|+|\delta_1|\leq 2\sqrt{\epsilon}$.     
\end{lemma}

\begin{proof}
The bounds on $|\beta|$ and $|\delta_b|$ are obtained by the same proof as Lemma \ref{beta}. 
So, we consider the bound of $|\delta_0|+|\delta_1|$. The fidelity requirement at $t_2$ gives us 
$||\pmb{\rho}(0)-\pmb{\rho}(1)||_{tr}\geq 2-4\epsilon$. 
Regard $\pmb{\rho}(0)$ and $\pmb{\rho}(1)$ as the points in the Bloch ball. 
By the triangle inequality, their distance is at most $(1-2|\delta_0|^2)+(1-2|\delta_1|^2)$. 
Thus, $|\delta_0|^2+|\delta_1|^2\leq 2\epsilon$. Then, it is easy to see 
that $|\delta_0|+|\delta_1|\leq 2\sqrt{\epsilon}$.    
\end{proof}

Similar to the proof of Theorem \ref{natural-bound}, we lead to two bounds on $\epsilon$ 
from the two paths. We first consider the path $s_1$-$t_1$. Let $C$ be the TP-CP map at $s_0$, 
and $M_1$ be the composite TP-CP map by the operations at $t_0$ and $t_1$. 
Take an arbitrary $|\psi\rangle$ and its orthogonal state $|\psi^\bot\rangle$. 
The fidelity requirement at $t_1$ gives us the condition 
\[
||M_1(C\otimes I)(\pmb{\rho}_\psi -\pmb{\rho}_{\psi^\bot})|\phi(b)\rangle\langle\phi(b)| ||_{tr}
\geq 2-4\epsilon. 
\]
Note that, letting $|\tilde{\phi}(b)\rangle=|\phi(b)_1\rangle|\phi(b)_2\rangle$, 
\[
|| |\phi(b)\rangle\langle\phi(b)| - |\tilde{\phi}(b)\rangle\langle\tilde{\phi}(b)| ||_{tr}
=2\sqrt{1-|\langle\phi(b)|\tilde{\phi}(b)\rangle|^2} =2|\delta_b|.
\]
Thus, by using the triangle inequality
\begin{align*}
&||C(\pmb{\rho}_\psi-\pmb{\rho}_{\psi^\bot})|\phi(b)_1\rangle\langle\phi(b)_1| ||_{tr}
=||(C\otimes I)(\pmb{\rho}_\psi-\pmb{\rho}_{\psi^\bot})|\tilde{\phi}(b)\rangle\langle\tilde{\phi}(b)| ||_{tr}\\
&\geq ||M_1(C\otimes I)(\pmb{\rho}_\psi-\pmb{\rho}_{\psi^\bot})|\tilde{\phi}(b)\rangle\langle\tilde{\phi}(b)| ||_{tr}\\
&\geq ||M_1(C\otimes I)(\pmb{\rho}_\psi-\pmb{\rho}_{\psi^\bot})|\phi(b)\rangle\langle\phi(b)| ||_{tr} 
- ||M_1(C\otimes I)(\pmb{\rho}_\psi-\pmb{\rho}_{\psi^\bot})
(|\tilde{\phi}(b)\rangle\langle\tilde{\phi}(b)|-|\phi(b)\rangle\langle\phi(b)|) ||_{tr}\\
&\geq 2-4\epsilon - ||\pmb{\rho}_{\psi}-\pmb{\rho}_{\psi^\bot}||_{tr}\cdot 2|\delta_b|\ 
\geq 2-4\epsilon - 4|\delta_b|.
\end{align*}
Let $C(b)$ be the TP-CP map that transforms $\pmb{\rho}$ to 
$C(\pmb{\rho}\otimes |\phi(b)_1\rangle\langle\phi(b)_1|)$. Then, we have 
\[
||C(b)(|\psi\rangle\langle\psi|- |\psi^\bot\rangle\langle\psi^\bot| )||_{tr}
\geq ||C(b)(\pmb{\rho}_{\psi} - \pmb{\rho}_{\psi^\bot} )||_{tr} 
\geq 2-4\epsilon -4|\delta_b|.
\]
This leads to the condition 
\begin{equation}\label{eq46}
||C(b)|\psi\rangle\langle\psi| ||_{tr}\geq 1-4\epsilon -4|\delta_b|
\end{equation}
since $|| C(b)|\psi^\bot\rangle\langle\psi^\bot| ||_{tr}\leq 1$. 
Recall that the Bloch sphere representation of $C(b)$ is written as the map: 
$\vec{r}\mapsto O_1(b)\Lambda(b)O_2(b)\vec{r} + \vec{d}(b)$ where $O_1(b)$, $O_2(b)$ 
are orthogonal matrices with determinant 1, and $\Lambda(b)$ is a diagonal matrix. 
Let $U(b)$ be the unitary operator whose Bloch sphere representation maps 
a Bloch vector $\vec{r}$ to $O_1(b)O_2(b)\vec{r}$. 
Now we take $|\psi\rangle$ such that $U(0)\pmb{\rho}_\psi=U(1)\pmb{\rho}_\psi$. 
Then, we evaluate $||(C(0)-C(1))\pmb{\rho'}_\psi||_{tr}$ 
with $\pmb{\rho'}_\psi=|\alpha|^2|\psi_1\rangle\langle\psi_1|$ as follows:
\begin{align*}
&||(C(0)-C(1))\pmb{\rho'}_\psi||_{tr}\\
&\leq ||(C(0)-U(0))\pmb{\rho'}_\psi||_{tr}+||U(0)(\pmb{\rho'}_\psi-\pmb{\rho}_\psi)||_{tr} 
+ ||U(1)(\pmb{\rho}_\psi-\pmb{\rho'}_\psi)||_{tr} +||(U(1)-C(1))\pmb{\rho'}_\psi||_{tr}\\
&= |\alpha|^2(||(C(0)-U(0))|\psi_1\rangle\langle\psi_1| ||_{tr}+ ||(C(1)-U(1))|\psi_1\rangle\langle\psi_1| ||_{tr})
+ 2||\pmb{\rho}_\psi-\pmb{\rho'}_{\psi}||_{tr}.
\end{align*}
Noting Ineq.(\ref{eq46}), $||U(b)|\psi_1\rangle\langle\psi_1| ||_{tr}=1$ 
and $||\pmb{\rho}_\psi-\pmb{\rho'}_{\psi}||_{tr} =|\beta|^2$, 
we obtain 
\begin{equation}\label{eq46-2}
||(C(0)-C(1))\pmb{\rho'}_\psi||_{tr}\leq 
|\alpha|^2(8\epsilon+4|\delta_0|+4|\delta_1|) +2|\beta|^2.
\end{equation}

Second, we consider the path $s_2$-$t_2$. Let $M_2$ be the composite map by the operations at $t_0$ and $t_2$. 
By the fidelity requirement at $t_2$, 
$||M_2(I\otimes C)|\xi_\psi\rangle\langle\xi_\psi|(\pmb{\rho}(0) -\pmb{\rho}(1))||_{tr}\geq 2-4\epsilon$. 
The left-hand side is at most
\begin{align*}
&||(I\otimes C)(|\psi_2\rangle\langle\psi_2|\otimes\pmb{\rho'})(\pmb{\rho}(0)-\pmb{\rho}(1))||_{tr}
+||(I\otimes C)(|\xi_\psi\rangle\langle\xi_\psi|-|\psi_2\rangle\langle\psi_2|\otimes\pmb{\rho'})
(\pmb{\rho}(0)-\pmb{\rho}(1))||_{tr}\\
&\leq ||C(\pmb{\rho'}(\pmb{\rho}(0)-\pmb{\rho}(1)))||_{tr} 
+ || |\xi_\psi\rangle\langle\xi_\psi|-|\psi_2\rangle\langle\psi_2|\otimes\pmb{\rho'} ||_{tr}
||\pmb{\rho}(0)-\pmb{\rho}(1)||_{tr}\\
&\leq ||(C'(0)-C'(1))\pmb{\rho'}||_{tr}+ 2|\beta|\sqrt{1-|\beta|^2/2}\times 2,
\end{align*}
where $C'(b)$ is the TP-CP map: $\pmb{\sigma}\mapsto C(\pmb{\sigma}\otimes\pmb{\rho}(b))$ 
and the last term of the right-hand side is obtained by the same calculation as Lemma \ref{just-calculation}. 
Since $\pmb{\rho}(b)=(1-2|\delta_b|^2)|\phi(b)_1\rangle\langle\phi(b)_1|+2|\delta_b|^2\cdot\frac{\pmb{I}}{2}$, 
$C'(b)$ is decomposed into $C'(b)=(1-2|\delta_b|^2)C(b)+2|\delta_b|^2 C_I$ with some TP-CP map $C_I$. 
Now we assume that $|\delta_0|^2\geq |\delta_1|^2$, which does not loss of generality. 
By the triangle inequality, 
\begin{align*}   
&||(C'(0)-C'(1))\pmb{\rho'}||_{tr}\\
&\leq ||(1-2|\delta_0|^2)(C(0)-C(1))\pmb{\rho'}||_{tr}+||(2|\delta_0|^2-2|\delta_1|^2) C(1)\pmb{\rho'}||_{tr} 
+||(2|\delta_0|^2-2|\delta_1|^2) C_I \pmb{\rho'}||_{tr}\\
&\leq (1-2|\delta_0|^2)||(C(0)-C(1))\pmb{\rho'}||_{tr}+4|\alpha|^2(|\delta_0|^2-|\delta_1|^2). 
\end{align*}
Thus, we have
\begin{equation}\label{eq46-3}
2-4\epsilon\leq 4|\beta|\sqrt{1-|\beta|^2/2}+
(1-2|\delta_0|^2)||(C(0)-C(1))\pmb{\rho'}||_{tr}+4|\alpha|^2(|\delta_0|^2-|\delta_1|^2). 
\end{equation}
By Ineqs.(\ref{eq46-2}) and (\ref{eq46-3}) from the two paths and Lemma \ref{beta-delta}, 
we obtain
\begin{align*}
2-4\epsilon &\leq 4|\beta|\sqrt{1-|\beta|^2/2}+
(1-2|\delta_0|^2)(|\alpha|^2(8\epsilon+4|\delta_0|+4|\delta_1|) +2|\beta|^2)
+4|\alpha|^2(|\delta_0|^2-|\delta_1|^2)\\
&\leq 4|\beta|+2|\beta|^2+4|\delta_0|^2+8\epsilon+4(|\delta_0|+|\delta_1|)\\
&\leq 4\sqrt{f(\epsilon)}+6f(\epsilon)+8\epsilon+8\sqrt{\epsilon},
\end{align*}
where $f(\epsilon)=\frac{1}{2}\left(\frac{3}{2}+\epsilon-\sqrt{\frac{9}{4}+\epsilon^2-5\epsilon}\right)$. 
Therefore, we have $1-4\sqrt{\epsilon}-6\epsilon\leq 2\sqrt{f(\epsilon)}+3f(\epsilon)$. 
The left-hand side is monotone decreasing on $\epsilon$ while the right-hand side 
is monotone increasing on $\epsilon$. By checking $\epsilon$ satisfying the inequality, 
we obtain $\epsilon > 0.017$. 
  
\subsection{Proof of Theorem \ref{thm-qubitbit}}
Here is the formal description of $XQC$.

\

\noindent
{\bf Protocol} $XQC$: Input $|\psi\rangle$ at $s_1$, and $b$ at $s_2$; 
Output $\pmb{\rho}_{out}^1$ at $t_1$, and $\mbox{Out}^2$ at $t_2$.

Step 1. $({\cal Q}_1,{\cal Q}_2)=UC(|\psi\rangle)$ at $s_1$, and ${\cal Q}_3={\cal Q}_4=|b\rangle$. 

Step 2. ${\cal Q}_5=X^b({\cal Q}_2)$ at $s_0$. 

Step 3. $({\cal Q}_6,{\cal Q}_7)=UC({\cal Q}_5)$ at $t_1$. 

Step 4 (Decoding at node $t_1$ and $t_2$). 
$\pmb{\rho}_{out}^1=X^b({\cal Q}_7)$, and $\mbox{Out}^2=0$ 
if $M[B_z]({\cal Q}_1)=M[B_z]({\cal Q}_6)$, $1$ otherwise.  
 
\

To average the fidelities at both sinks, implement $XQC$ 
with probability $3/4$ and replace Step 1 by the following operation 
with probability $1/4$: send a bit $r$ uniformly at random from $s_1$ 
to $s_0$ and $t_2$. 

Now we show that $XQC$ achieves fidelity $13/18$ and $11/18$ at $t_1$ and $t_2$, 
respectively. (The analysis of fidelity $2/3$ by averaging is omitted since it 
is rather simpler.) 
First, we show that the fidelity at $t_1$ is $13/18$. For this purpose, consider the 
path from $s_1$ to $t_1$ in Fig.\ \ref{prot1}. Note that along the path, 
the operations at nodes $s_1$ and $t_0$ are $2/3$-shrinking, and the operation at $s_0$ is $X^b$. 
They are commutative, and hence the composite operation on the system ${\cal Q}_2$ 
along this path can be considered as a $4/9$-shrinking map followed by $X^b$. 
Since $X^b$ is applied at $t_1$ again, the final state at $t_1$ is $\frac{4}{9}|\psi\rangle\langle\psi| 
+ \frac{5}{9}\cdot\frac{\pmb{I}}{2}$. Hence, the fidelity is $F(\pmb{\rho}_{out}^1,|\psi\rangle) = 13/18$. 

Second, we show that the probability of recovering $b$ at node $t_2$ is $11/18$. 
For this purpose, we have to consider the path from $s_2$ to $t_2$ in Fig.\ \ref{prot1} 
and the entangled state of ${\cal Q}_1\otimes {\cal Q}_2$. 
Particularly, the cloning at $t_0$, which only operates on the later half of the system, 
can be regarded as $p$-shrinking map on ${\cal Q}_2$ in the following sense.

\begin{lemma}\label{shrink2}
Let $\pmb{\rho}_{12}$ be a two-qubit state and  $\pmb{\rho}_{12}'$ be the two-qubit state after applying 
a $p$-shrinking map on the second qubit of $\pmb{\rho}_{12}$. Then, 
$$\pmb{\rho}_{12}' = p\pmb{\rho}_{12} + 
(1-p)\left(\mathrm{Tr}_2(\pmb{\rho}_{12}) \otimes \frac{\pmb{I}}{2}\right).$$
\end{lemma}

\begin{proof}%{Lemma \ref{shrink2}}
Consider the cases of the $1$-shrinking and $0$-shrinking maps on the
second qubit of $\pmb{\rho}_{12}$. In those cases, $\pmb{\rho}_{12}'$
is mapped to $\pmb{\rho}_{12}$ and $\left(\mbox{Tr}_2(\pmb{\rho}_{12}) \otimes
\frac{\pmb{I}}{2}\right)$, respectively. (For $p = 0$, this is verified from 
the facts that $\mbox{Tr}_1(\pmb{\rho}_{12}') = \frac{\pmb{I}}{2}$ and
$\mbox{Tr}_2(\pmb{\rho}_{12}') = \mbox{Tr}_2(\pmb{\rho}_{12})$ for any
$\pmb{\rho}_{12}$.) For $0 < p < 1$, it is easy to see that
$\pmb{\rho}_{12}'$ can be described as a linear combination 
of the above two cases from the linearity of quantum operation.
\end{proof}

Now, it is easy to compute the success probability of recovering $b$ at 
node $t_2$ using the above lemma for $p = 2/3$. Let $\pmb{\rho}_{12}$ and 
$\pmb{\rho}'_{12}$ be the two-qubit state on ${\cal Q}_1\otimes{\cal Q}_2$ 
and on ${\cal Q}_1\otimes{\cal Q}_6$, respectively. We can check that    
$\langle 00| \pmb{\rho}_{12} |00\rangle + \langle 11| \pmb{\rho}_{12} 
|11\rangle= \frac{2}{3}$, and 
%\langle 01| \pmb{\rho}_{12} |01\rangle + \langle 10| \pmb{\rho}_{12} 
%|10\rangle &=& \frac{7}{18},\\
$\langle 00| \mbox{Tr}_2(\pmb{\rho}_{12})\otimes \frac{\pmb{I}}{2} |00\rangle + 
\langle 11| \mbox{Tr}_2(\pmb{\rho}_{12})\otimes \frac{\pmb{I}}{2} |11\rangle= \frac{1}{2}$. 
Note that when $b = 0$, $\pmb{\rho}_{12}$ does not change at $s_0$. 
By Lemma \ref{shrink2}, the probability of recovering $b = 0$ at node $t_2$ is 
$$\langle 00| \pmb{\rho}_{12}' |00\rangle + \langle 11| \pmb{\rho}_{12}' 
|11\rangle = \frac{2}{3}\cdot\frac{2}{3} + \frac{1}{3}\cdot\frac{1}{2} = 
\frac{11}{18}.$$  
On the other hand when $b = 1$, $(I \otimes X)$ is applied to 
$\pmb{\rho}_{12}$ at $s_0$ and therefore the probability of error, i.e., 
that of obtaining $b = 0$ at node $t_2$ is 
$$\langle 00| (I\otimes X) \pmb{\rho}_{12}' (I\otimes X)|00\rangle + 
\langle 11| (I\otimes X) \pmb{\rho}_{12}' 
(I\otimes X) |11\rangle = \langle 01| \pmb{\rho}_{12}' |01\rangle + \langle 10| \pmb{\rho}_{12}' 
|10\rangle = 1 - \frac{11}{18}= \frac{7}{18}$$  
since $I\otimes X$ is commutative with the $p$-shrinking map on the second qubit.
Thus we obtain the fidelity $11/18$ at $t_2$. 

\subsection{Proof of Theorem \ref{thm-2bits}}
First, we recall the quantum random access (QRA) coding by Ambainis et al. \cite{ANTV02}.  
An {\em $(n,m,p)$-QRA coding} is a function that maps $n$-bit strings $x \in \{0,1\}^n$ 
to $m$-qubit states $\pmb{\rho}_x$ satisfying the following: For every $i\in\{1,2,\ldots,n\}$ 
there exists a POVM $E^{i} = \{E_0^{i},E_1^{i}\}$ such that 
$\mathrm{Tr}(E_{x_i}^{i}\pmb{\rho}_x)\geq p$ for all $x\in\{0,1\}^n$, where $x_i$ is the $i$-th bit of $x$. 
If the $m$-qubit states are classical, the coding is called an {\em $(n,m,p)$-classical random access coding}. 
In \cite{ANTV02}, an $(2,1,0.85)$-QRA coding is given by the following protocol. 
\begin{quote}
{Let $|\varphi(00)\rangle=\cos(\pi/8)|0\rangle+\sin(\pi/8)|1\rangle$, 
$|\varphi(10)\rangle=\cos(3\pi/8)|0\rangle+\sin(3\pi/8)|1\rangle$, 
$|\varphi(11)\rangle=\cos(5\pi/8)|0\rangle+\sin(5\pi/8)|1\rangle$, 
and $|\varphi(01)\rangle=\cos(7\pi/8)|0\rangle+\sin(7\pi/8)|1\rangle$ be the one-qubit state used 
when the source $x\in\{0,1\}^2$ is respectively $00$, $10$, $11$, and $01$. %(See Fig.~\ref{twotoone}.)  
The first bit of $x$ is obtained by measuring in the basis $B_z$, while the second one by measuring in the basis $B_x$.}
\end{quote}
In fact, the success probability of the above protocol is $\cos^2(\pi/8)\approx 0.85$. 
On the contrary, it was also shown that any $(2,1,p)$-classical random access coding should satisfy $p\leq 1/2$. 

A map that transforms an arbitrary equatorial state (i.e., the one-qubit state whose amplitudes are real) 
$|\psi\rangle\langle\psi|$ to $p|\psi\rangle\langle\psi|+(1-p)\frac{\pmb{I}}{2}$ is called 
a $p$-shrinking map on equatorial qubits. 
The following lemma is verified from the transformation of the phase-covariant cloning machine 
\cite{Bru00,FMWI02}. (The term ``the phase-covariant copy'' is defined similarly as 
the universal copy.) Fortunately, for our purpose the detail of the cloning machine except 
the fact that it is a $1/\sqrt{2}$-shrinking map is not needed. 

\begin{lemma}\label{shrink-phase}
The phase-covariant copy is $1/\sqrt{2}$-shrinking map on equatorial qubits.  
\end{lemma}

Finally we introduce the 2D measurement. This measurement is defined 
by the POVM $\{\frac{1}{2}|\varphi(z_1z_2)\rangle \mid z_1z_2\in \{0,1\}^2 \}$, denoted by $MM_2$. 
Its intuition is to do the two projective measurements in the bases $\{|\varphi(00)\rangle,|\varphi(11)\rangle\}$ 
and $\{|\varphi(01)\rangle,|\varphi(10)\rangle\}$ with probability $1/2$ for each. 
Notice that we estimate the QRA coding state from $s_2$ correctly if we choose the right one of the two bases. 

The detailed description of $X2C2C$ is as follows. The term $({\cal Q},{\cal Q}')=PC({\cal Q}'')$ 
means that ${\cal Q}$ and ${\cal Q}'$ are the two copies output by the phase-covariant quantum cloning machine 
when ${\cal Q}''$ is given as the input. 

\

\noindent 
{\bf Protocol} $X2C2C$: Input $x_1x_2$ at $s_1$, $y_1y_2$ at $s_2$; 
Output $\mbox{Out}^1$ at $t_1$, $\mbox{Out}^2$ at $t_2$.  

Step 1. ${\cal Q}_1=|\varphi(x_1x_2)\rangle$, ${\cal Q}_2=|\varphi(x_1x_2)\rangle$, 
${\cal Q}_3=|\varphi(y_1y_2)\rangle$, and ${\cal Q}_4=|\varphi(y_1y_2)\rangle$. 

Step 2. ${\cal Q}_5= GR({\cal Q}_2,MM_2({\cal Q}_3))$ at $s_0$. 

Step 3. $({\cal Q}_6,{\cal Q}_7)=PC({\cal Q}_5)$ at $t_0$. 

Step 4 (Decoding the $j$-th bit at $t_1$ and $t_2$). 
$\mbox{Out}^1=M[B_z]({\cal Q}_4)\oplus M[B_z]({\cal Q}_7)$ if $j=1$, 
and $M[B_x]({\cal Q}_4)\oplus M[B_x]({\cal Q}_7)$ if $j=2$.  
$\mbox{Out}^2=M[B_z]({\cal Q}_1)\oplus M[B_z]({\cal Q}_6)$ if $j=1$, 
and $M[B_x]({\cal Q}_1)\oplus M[B_x]({\cal Q}_6)$ if $j=2$. 

\

Here, we analyze the success probability of $X2C2C$. By the definition of $MM_2$ the following lemma 
is immediate.
\begin{lemma}\label{mm2}
Given a QRA coding state $|\varphi(y_1y_2)\rangle$, the probability that $z_1z_2$ is obtained 
by $MM_2$ is $1/2$ if $z_1z_2=y_1y_2$, $1/4$ if $z_1z_2=\bar{y}_1y_2$ or $y_1\bar{y}_2$, 
and $0$ if $z_1z_2=\bar{y}_1\bar{y}_2$. Here, $\bar{b}$ denotes the negation of a bit $b$.
\end{lemma}
For simplicity, let us assume that $y_1y_2 = 00$ (the analysis of the other cases are similar). 
By Lemma \ref{mm2}, the state of ${\cal Q}_2\otimes{\cal Q}_3$ after $MM_2$ is
\[
|\varphi(x_1x_2)\rangle\langle\varphi(x_1x_2)|\otimes
\left(\frac{1}{2}|00\rangle\langle 00|+\frac{1}{4}(|10\rangle\langle 10|+|01\rangle\langle 01|)
\right).
\]
Note that if $\tilde{y}_1\tilde{y}_2$ is obtained by $MM_2$, the state $|\varphi(x_1x_2)\rangle$ 
from $s_1$ is transformed into $|\varphi(x_1\oplus \tilde{y}_1,x_2\oplus \tilde{y}_2)\rangle$ by $GR$. 
Thus, the state of ${\cal Q}_5$ is 
\[
\frac{1}{2}|\varphi(x_1x_2)\rangle\langle \varphi(x_1x_2)|
+\frac{1}{4}( |\varphi ( \bar{x}_1x_2)\rangle\langle \varphi (\bar{x}_1x_2)|
+|\varphi (x_1\bar{x}_2)\rangle\langle \varphi (x_1\bar{x}_2)| )
\]
By Lemma \ref{shrink-phase}, the state of ${\cal Q}_6$ (or ${\cal Q}_7$) is 
\[
\frac{1}{2\sqrt{2}}|\varphi(x_1x_2)\rangle\langle \varphi(x_1x_2)|
+\frac{1}{4\sqrt{2}}(|\varphi(\bar{x}_1x_2)\rangle\langle \varphi(\bar{x}_1x_2)|
+|\varphi(x_1\bar{x}_2)\rangle\langle \varphi(x_1\bar{x}_2)|)
+\left(1-\frac{1}{\sqrt{2}}\right)\frac{\pmb{I}}{2}.
\]
Now we check the success probability of decoding the first bit of $y_1y_2=00$, i.e., $0$ 
at $t_2$ (the other cases are similarly checked). As seen from the state of ${\cal Q}_6$, 
the success probability of decoding the first bit of $x_1x_2\oplus y_1y_2$, i.e., $x_1$ is 
\[
\frac{1}{2\sqrt{2}}\cdot \cos^2\frac{\pi}{8}+\frac{1}{4\sqrt{2}}(\cos^2\frac{\pi}{8}+\sin^2\frac{\pi}{8})
+\frac{1}{2} \left(1-\frac{1}{\sqrt{2}}\right),
\]
which is $5/8$. On the contrary, the success probability of decoding the first bit $x_1$ from the state of ${\cal Q}_1$ 
is $\cos^2\frac{\pi}{8}$. Thus, the success probability of decoding the first bit of $y_1y_2$ at $t_2$ 
is $\cos^2\frac{\pi}{8}\cdot\frac{5}{8}+\sin^2\frac{\pi}{8}\cdot\frac{3}{8}=1/2+\sqrt{2}/16$ as claimed.

\subsection{Proof of Theorem \ref{x3c3c}}
For the proof of Theorem \ref{x3c3c}, we need the following two primitives. 

\begin{figure}\begin{center}
\includegraphics*[width=7cm]{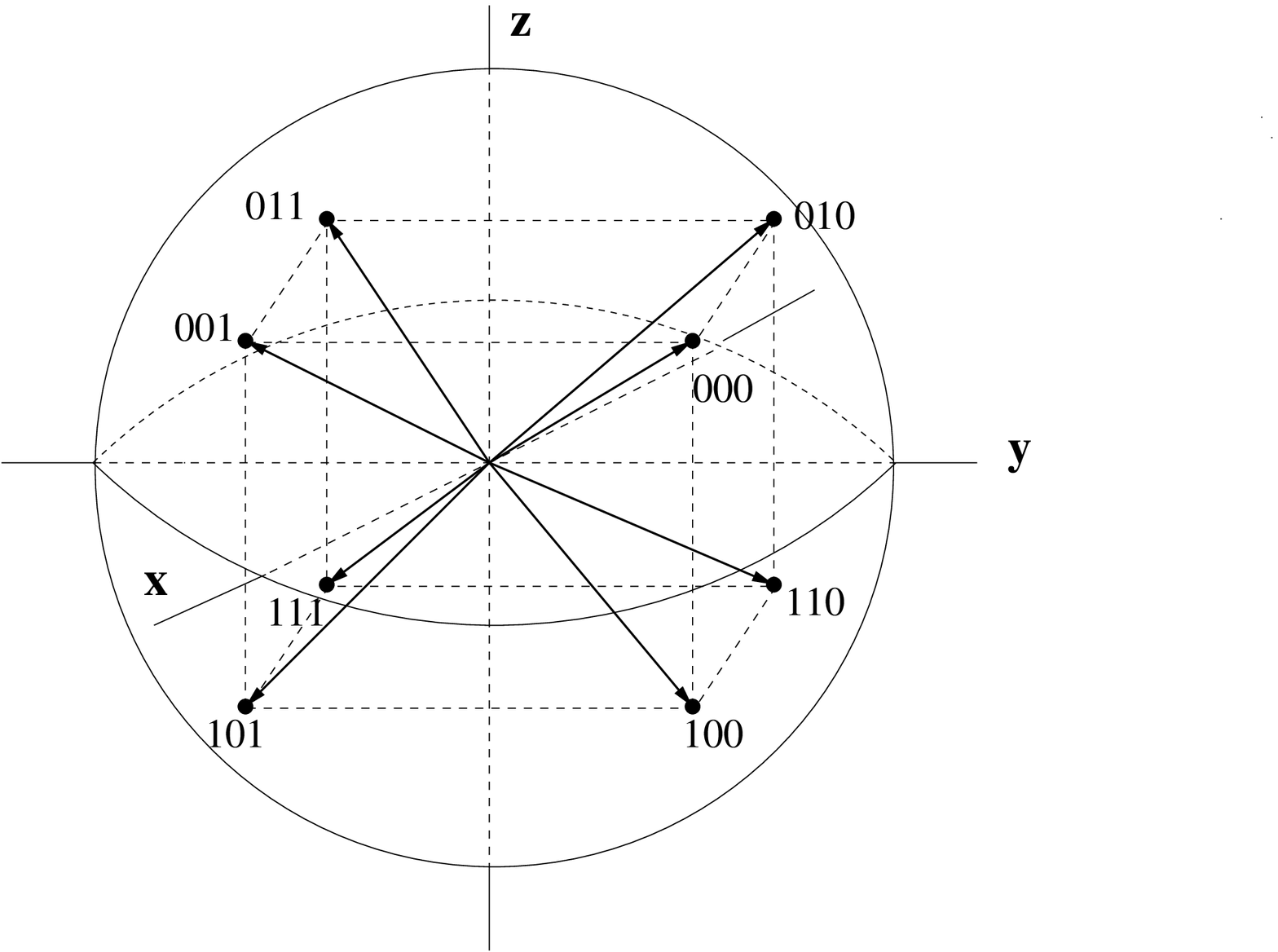}
\caption{$(3,1,0.79)$-QRA coding in the Bloch sphere representation}\label{threetoone}
\end{center}
\end{figure}

{\bf 3D Measurement (MM$_3$).} 
The {\em 3D measurement}, denoted by $MM_3$, is defined by the POVM described 
by $\{\frac{1}{4}|\varphi(z_1z_2z_3)\rangle\langle\varphi(z_1z_2z_3)|\mid z_1z_2z_3\in\{0,1\}^3 \}$, 
where $|\varphi(z_1z_2z_3)\rangle$ is the $(3,1,0.79)$-QRA coding state of $z_1z_2z_3$ (Fig.\ \ref{threetoone}). 

{\bf Approximated Group Operation (AG).} Before its definition, we introduce the operation $\mathrm{Inv}$, 
which maps a pure state $|\psi\rangle=a|0\rangle+b|1\rangle$ to its ``opposite'' state 
$b^*|0\rangle-a^*|1\rangle$ in the Bloch sphere (see e.g., \cite{BHW,GP99}). 
Formally, it is defined as follows: For any mixed state 
$\pmb{\rho}=\left(\begin{array}{ll}
u & v \\w & x
\end{array}
\right)$, 
$\mathrm{Inv}\pmb{\rho} =
\left(\begin{array}{ll}
x & -v \\ -w & u
\end{array}\right)$. 
Note that the set of maps $\pmb{\rho}\mapsto W\pmb{\rho}W^\dagger$  
($W\in\{I,X,Y,Z,\mathrm{Inv},\mathrm{Inv}X,\mathrm{Inv}Y,\mathrm{Inv}Z\}$) 
is an abelian group. Because $\mathrm{Inv}$ is not a TP-CP map, 
we introduce its approximation $\mathrm{Inv}'$, which maps $\pmb{\rho}$ 
to $\mathrm{Inv}'\pmb{\rho} =\frac{1}{3}\mathrm{Inv}\pmb{\rho} + \frac{2}{3}\cdot\frac{\pmb{I}}{2}$. 
We can check that $\mathrm{Inv}'$ is a TP-CP map. 
%%|0>|00> -> (|00>+|11>)|1>+|01>|0>; |1>|00> -> (|00>-|11>)|0>-|01>|1>
The {\em approximated group operation under} 
a three-bit string $r_1r_2r_3$, denoted by $AG(\pmb{\rho},r_1r_2r_3)$, 
is a transformation defined by $AG(\pmb{\rho},000)=\pmb{\rho}$, $AG(\pmb{\rho},011)=Z\pmb{\rho}$, 
$AG(\pmb{\rho},101)=X\pmb{\rho}$, $AG(\pmb{\rho},110)=Y\pmb{\rho}$ and, for any $r_1r_2r_3\in\{001,010,100,111\}$, 
$AG(\pmb{\rho},r_1r_2r_3)=\mathrm{Inv}'AG(\pmb{\rho},\bar{r}_1\bar{r}_2\bar{r}_3)$. 

The description of the protocol $X3C3C$ is as follows. 

\

\noindent
{\bf Protocol $X3C3C$}: Input $x_1x_2x_3$ at $s_1$, $y_1y_2y_3$ at $s_2$; Output $\mathrm{Out}^1$ at $t_1$,  $\mathrm{Out}^2$ at $t_2$. 

Step 1. ${\cal Q}_1=|\varphi(x_1x_2x_3)\rangle$, ${\cal Q}_2=|\varphi(x_1x_2x_3)\rangle$, 
${\cal Q}_3=|\varphi(y_1y_2y_3)\rangle$, and ${\cal Q}_4=|\varphi(y_1y_2y_3)\rangle$. 

Step 2. ${\cal Q}_5= AG({\cal Q}_2,MM_3({\cal Q}_3))$ at $s_0$. 

Step 3. $({\cal Q}_6,{\cal Q}_7)= UC({\cal Q}_5)$ at $t_0$. 

Step 4 (Decoding the $j$-th bit at $t_1$ and $t_2$). 
$\mbox{Out}^1=M[B_z]({\cal Q}_4)\oplus M[B_z]({\cal Q}_7)$ if $j=1$, 
$M[B_x]({\cal Q}_4)\oplus M[B_x]({\cal Q}_7)$ if $j=2$, and 
$M[B_y]({\cal Q}_4)\oplus M[B_y]({\cal Q}_7)$ if $j=3$.  
$\mbox{Out}^2=M[B_z]({\cal Q}_1)\oplus M[B_z]({\cal Q}_6)$ if $j=1$, 
$M[B_x]({\cal Q}_1)\oplus M[B_x]({\cal Q}_6)$ if $j=2$, and 
$M[B_y]({\cal Q}_1)\oplus M[B_y]({\cal Q}_6)$ if $j=3$. 

\

Now, we analyze $X3C3C$, which is similar to $X2C2C$. 
By definition of POVM $MM_3$, we can easily check the following lemma. 

\begin{lemma}\label{povm-mm3}
Given a QRA coding state $|\varphi(y_1y_2y_3)\rangle$, the probability that $z_1z_2z_3$ 
is obtained by $MM_3$ is 
$$
\left\{\begin{array}{ll}
1/4 & \mbox{if }z_1z_2z_3=y_1y_2y_3,\\ 
1/6 & \mbox{if }z_1z_2z_3=\bar{y}_1y_2y_3,\ y_1\bar{y}_2y_3,\ y_1y_2\bar{y}_3,\\
1/12 & \mbox{if }z_1z_2z_3=\bar{y}_1\bar{y}_2y_3,\ \bar{y}_1y_2\bar{y}_3,\ y_1\bar{y}_2\bar{y}_3,\\
0   & \mbox{if }z_1z_2z_3=\bar{y}_1\bar{y}_2\bar{y}_3
\end{array}
\right.
$$
\end{lemma}
Henceforth, for simplicity of descriptions we only consider the case $y_1y_2y_3=000$. 
By Lemma \ref{povm-mm3}, the state of ${\cal Q}_2\otimes{\cal Q}_3$ after $MM_3$ is 
\begin{align*}
|\varphi(x_1x_2x_3)\rangle & \langle\varphi(x_1x_2x_3)|
\otimes
\left(
\frac{1}{4}|000\rangle\langle 000|
+\frac{1}{6}(|100\rangle\langle 100|+|010\rangle\langle 010|
\right.\\
&\left. +|001\rangle\langle 001|)
 +\frac{1}{12}(|110\rangle\langle 110|
+|101\rangle\langle 101|+|011\rangle\langle 011|)
\right).
\end{align*}
Noting that $\mathrm{Inv}'(|\varphi(z_1z_2z_3)\rangle\langle\varphi(z_1z_2z_3)|)=\frac{1}{3}
|\varphi(\bar{z}_1\bar{z}_2\bar{z}_3)\rangle\langle\varphi(\bar{z}_1\bar{z}_2\bar{z}_3)|
+\frac{2}{3}\cdot\frac{\pmb{I}}{2}$, the state of ${\cal Q}_5$ is 
\begin{align*}
&\frac{1}{4}|\varphi(x_1x_2x_3)\rangle\langle \varphi(x_1x_2x_3)|\\
&+\frac{1}{18}(|\varphi(\bar{x}_1x_2x_3)\rangle\langle \varphi(\bar{x}_1x_2x_3)|
+|\varphi(x_1\bar{x}_2x_3)\rangle\langle \varphi(x_1\bar{x}_2x_3)|
+|\varphi(x_1x_2\bar{x}_3)\rangle\langle \varphi(x_1x_2\bar{x}_3)|)\\
&+\frac{1}{12}(
|\varphi(\bar{x}_1\bar{x}_2x_3)\rangle\langle \varphi(\bar{x}_1\bar{x}_2x_3)|
+|\varphi(\bar{x}_1x_2\bar{x}_3)\rangle\langle \varphi(\bar{x}_1x_2\bar{x}_3)|
+|\varphi(x_1\bar{x}_2\bar{x}_3)\rangle\langle \varphi(x_1\bar{x}_2\bar{x}_3)|
)+\frac{2}{3}\cdot\frac{1}{6}\cdot 3\cdot\frac{\pmb{I}}{2}.
\end{align*}
By Lemma \ref{shrink}, the state of ${\cal Q}_6$ (or ${\cal Q}_7)$) is 
\begin{align*}
&\frac{1}{6}|\varphi(x_1x_2x_3)\rangle\langle \varphi(x_1x_2x_3)|\\
&+\frac{1}{27}(|\varphi(\bar{x}_1x_2x_3)\rangle\langle \varphi(\bar{x}_1x_2x_3)|
+|\varphi(x_1\bar{x}_2x_3)\rangle\langle \varphi(x_1\bar{x}_2x_3)|
+|\varphi(x_1x_2\bar{x}_3)\rangle\langle \varphi(x_1x_2\bar{x}_3)|)\\
&+\frac{1}{18}(
|\varphi(\bar{x}_1\bar{x}_2x_3)\rangle\langle \varphi(\bar{x}_1\bar{x}_2x_3)|
+|\varphi(\bar{x}_1x_2\bar{x}_3)\rangle\langle \varphi(\bar{x}_1x_2\bar{x}_3)|
+|\varphi(x_1\bar{x}_2\bar{x}_3)\rangle\langle \varphi(x_1\bar{x}_2\bar{x}_3)|
)+\frac{5}{9}\cdot\frac{\pmb{I}}{2},
\end{align*}
where the last term is obtained by adding $\frac{1}{3}\cdot\frac{1}{4}\cdot\frac{\pmb{I}}{2}$, 
$\frac{1}{3}\cdot\frac{1}{18}\cdot 3\cdot\frac{\pmb{I}}{2}$, 
$\frac{1}{3}\cdot\frac{1}{12}\cdot 3\cdot\frac{\pmb{I}}{2}$, 
and $\frac{2}{3}\cdot\frac{1}{6}\cdot 3\cdot\frac{\pmb{I}}{2}$. 
Now we check the success probability of decoding the first bit of 
$y_1y_2y_3=000$, i.e., $0$. (The other cases are similarly checked.) 
As seen from the state of ${\cal Q}_6$, the success probability of decoding the first bit 
of $x_1x_2x_3\oplus y_1y_2y_3$, i.e., $x_1$ is 
\begin{align*}
&\frac{1}{6}\left(\frac{1}{2}+\frac{\sqrt{3}}{6}\right)
+\frac{1}{27}\left(
2\cdot\left(\frac{1}{2}+\frac{\sqrt{3}}{6}\right)+\left(\frac{1}{2}-\frac{\sqrt{3}}{6}\right) 
\right)\\
&+\frac{1}{18}\left(
2\cdot\left(\frac{1}{2}-\frac{\sqrt{3}}{6}\right)+\left(\frac{1}{2}+\frac{\sqrt{3}}{6}\right) 
\right)+\frac{5}{9}\cdot\frac{1}{2}\ 
=\frac{1}{2}+\frac{2\sqrt{3}}{81}.
\end{align*}
On the contrary, the success probability of decoding the first bit $x_1$ from the state of ${\cal Q}_1$ 
is $1/2+\sqrt{3}/6$. Thus, the success probability of decoding the first bit of 
$y_1y_2y_3$ at sink $t_2$ is $(1/2+\sqrt{3}/6)(1/2+2\sqrt{3}/81)+(1/2-\sqrt{3}/6)(1/2-2\sqrt{3}/81)=1/2+2/81$ 
as desired.

\end{document}